\documentclass[a4paper,10pt]{article}
\usepackage[round]{natbib}
\usepackage[english]{babel}
\usepackage{setspace}
\usepackage[utf8x]{inputenc}
\usepackage{graphicx} 
\usepackage[left=2cm,right=2cm,top=2cm,bottom=2cm]{geometry}
\usepackage{textcomp}
\usepackage[T1]{fontenc}
\usepackage{amssymb}
\usepackage{xcolor}
\usepackage{lineno}
\usepackage{authblk}

\usepackage{ulem}
\usepackage{appendix}
\usepackage{hyperref}
\usepackage{float}

\title{
\large
\textbf{Cloud-convection feedback in brown dwarfs atmosphere}}

\author[1]{Maxence Lef{\`e}vre}
\author[1]{Xianyu Tan}
\author[2]{Elspeth K. H. Lee}
\author[1]{R. T. Pierrehumbert}
\affil[1]{Department of Physics (Atmospheric, Oceanic and Planetary Physics), University of Oxford, Parks Rd, Oxford, OX1 3PU, UK.}
\affil[2]{Center for Space and Habitability, University of Bern, Gesellschaftsstrasse 6, CH-3012 Bern, Switzerland}

\begin{document}

\maketitle
\newpage

\section*{Abstract}
Numerous observational evidence has suggested the presence of active meteorology in the atmospheres of brown dwarfs. A near-infrared brightness variability has been observed. Clouds have a major role in shaping the thermal structure and spectral properties of these atmospheres. The mechanism of such variability is still unclear and both 1D and global circulation model cannot fully study this topics due to resolution. In this study, a convective resolving model is coupled to grey-band radiative transfer in order to study the coupling between the convective atmosphere and the variability of clouds over a large temperature range with a domain of several hundreds of kilometers. Six types of clouds are considered, with microphysics including settling. The clouds are radiatively active using Rosseland mean coefficient. Radiative cloud feedback can drive spontaneous atmospheric variability in both temperature and cloud structure, as modeled for the first time in three dimensions. Silicate clouds have the most effect of the thermal structure with the generation of a secondary convective layer in some cases, depending on the assumed particle size. Iron and Aluminum clouds also have a substantial impact on the atmosphere. Thermal spectra were computed, and we find the strongest effect of clouds is the smoothing of spectral features at optical wavelengths. Compared to observed L and T dwarfs on color-magnitude diagram, the simulated atmospheres are redder for most of the cases. The simulations with the presence of cloud holes are closer to the observations.

\section{Introduction}

Since the first two confirmations of the observations of brown dwarfs \citep{Naka95,Rebo96}, about 2000 similar objects have been observed, the vast majority of them free-floating. These objects are classified in three types: L, T and Y dwarfs in decreasing temperature range. The L dwarfs are cooler than the M dwarf but exhibit similarities in photospheric chemical composition, containing alkali lines (K, Na), metal hybrids (FeH), oxides (TiO, VO) and water. The so-called L/T transition occurs at effective temperatures around 1100–1400 K. The L dwarfs appear then red in the color–magnitude diagram, with CO absorption, while T dwarfs appear bluer with stronger CH$_4$ absorption features \citep{Kirk05,Cush14}. The chemical change is attributed to the cooling of the atmosphere. The L/T transition is presumably linked to the formation of iron and silicate clouds in the photosphere of L dwarfs, shifting them redder in the color–magnitude diagram \citep{Tsuj96,Acke01,Alla01,Burr06}. The Y dwarfs are colder and show H$_2$O and CH$_4$ absorption bands in the near-infrared, as well as water clouds features \citep{Morl14}.
\bigbreak
The variability of the light-curve in the IR is thought to be caused by inhomogeneous surface brightness due to rotational modulation \citep{Radi12,Apai13,Kara16}, evolving over a few rotation periods \citep{Arti09,Radi12,Bill13,Gill13,Metc15,Apai17}, suggesting a rapid change of the surface features. Horizontal variation of the cloud and temperature structure \citep{Radi12,Apai13,Buen15,Kara16} could also play a role in this variability, but the mechanisms controlling clouds dynamics are still under investigations.
\bigbreak
The effect of clouds has been studied using 1D radiative–convective models, with sophisticated models computing the cloud self-consistently with microphysics and parametric cloud models using fixed parameters to describe the cloud particle distribution. For the first category, there is the model Drift-Phoenix \citep{Hell08,Witt09} with a full microphysics scheme with cloud formation on TiO$_2$; \cite{Gao18} with condensation of KCl clouds; BT-Settl \citep{Alla12} which includes 40 condensable species; Exo-REM \citep{Char18} using iron, silicate and sulfide and alkali salt clouds and \cite{Tan19} using MgSiO$_3$ clouds. For the second category, there is \cite{Tsuj02} includes corundum, iron and silicate clouds; \cite{Acke01,Marl10,Morl12,Morl14} which includes iron, silicate, sulfide, salt, and water clouds, and \cite{Burr06,Madh11} which includes iron and silicate clouds.
\bigbreak
The dynamics of such objects have been studied with global circulation models (GCM), \cite{Show13} presented the first global models of interior convection, showing the importance of rotation in the dynamics. \cite{Zhan14} performed global simulations using a one-and-a-half layer shallow-water model. \cite{Tan17} studied the effect of latent heating associated with condensation of silicates and iron on the dynamical circulation using an idealized 3D GCM. \cite{Show19} parameterized interactions between convection and the stratified layers as isotropic, stochastic temperature perturbations in the GCM and explored the effect on zonal jet formation. \cite{Tan21a,Tan21b} studied the effects of enstatite cloud radiative feedback on the global circulation. Several models studied the dynamics of the irradiated brown dwarfs \citep{Lee20,Tan20}. 
\bigbreak
The GCM and 1D model have in common that they do not resolve turbulence, and therefore need to use parametrization for the convective activity. The vertical transport is represented by a vertical eddy diffusivity coefficient K$_{zz}$ that varies depending on the hypothesis \citep{Zhan20}. Very few studies have been conducted using a local non-hydrostatic approach, resolving the turbulence and thus not needing the K$_{zz}$ coefficient. \cite{Frey10} using a 2D fully compressible equations modeled Mg$_2$SiO$_4$ clouds for temperature between 900 and 2800~K, founding that convectively excited gravity waves are important for vertical mixing in the atmospheres. \cite{Bord18} performed local both 2D and 3D fully compressible simulations studying the turbulent mixing of CO and NH$_3$ vapor without clouds. A new scale height was developed, derived from the chemical scale height, to predict quench points and perform 1D chemical kinetics modeling. \cite{Trem21} studied radiative Rayleigh-Taylor instabilities that could influence the cloud cover.
\bigbreak
In this study, we use a 3D local non-hydrostatic dynamical core coupled to a grey-band radiative transfer and an idealized microphysical scheme to study the vertical turbulent mixing and its impact on the convective layer for hot brown dwarfs for different cloud composition, and compare the results with 1D modeling and observations. 
\bigbreak
The paper is organized as follows. Our Cloud Resolving Model (CRM) is described in Section~\ref{Sec:model}. In Section~\ref{Sec:free}, the clear-sky simulations are presented, and the impact of clouds on the emission spectra is discussed in Sections \ref{Sec:cloud}. Our conclusions are summarized in Section~\ref{Sec:conc}.

\section{Model Description}
\label{Sec:model}

\subsection{Dynamical core}
\label{Sec:model1}

The CRM is using the non-hydrostatic, compressible dynamical core CM1 version 19 \citep{Brya02} in Large-Eddy Simulation mode. The conservation of mass, momentum, and entropy is ensured by an explicitly conservative flux-form formulation of the fundamental equations, based on mass-coupled meteorological variables (winds and potential temperature) using third-order Runge–Kutta time differencing and fifth-order spatial derivative. A subgrid-scale prognostic turbulent kinetic energy closure \citep{Dear80} is used to parameterize the turbulent mixing by unresolved small-scale eddies, following the strategy adopted by \cite{Kang11,Wang14,Mark16,Shi18}. This study is the first one to apply the CM1 dynamical core to a hydrogen atmosphere and the second applications of non-Earth atmosphere \citep{Tan21}.

\subsection{Radiative Transfer}
\label{Sec:model2}

The radiative transfer used in this study is a plane-parallel, two-stream approximation, with a gray atmosphere with a single broad thermal band for simplicity and computational efficiency. The radiative transfer equations are solved using the numeral package TWOSTR \citep{Kyll95}. Absorbing, emitting, and multiple-scattering atmosphere are taken into account. A frequency-averaged gas opacity is used for the background model atmosphere, the Rosseland-mean opacity $\kappa_{R,g}$ from \cite{Free14} for 12 species (CH$_4$, CO, CO$_2$, CrH, FeH, H$_2$O, H$_2$S, H$_2$, NH$_3$, PH$_3$, TiO, VO) and H$_2$ CIA, over range of metallicity (for a mass fraction between 0 and 1.7). The interactions of cloud particles with radiation by absorption and scattering are parameterized by an extinction coefficient Q$_{ext}$, a scattering coefficient Q$_{scat}$, and an asymmetry parameter g. The total cloud extinction opacity $\kappa_{R,ext}$ is averaged over wavelength

\begin{equation}
    \frac{1}{\kappa_{R,ext}}= \frac{\int_{0}^{\infty}\frac{1}{\kappa_{ext}(\lambda)} \frac{dB_\lambda}{dT}d\lambda}{\int_{0}^{\infty} \frac{dB_\lambda}{dT}d\lambda}
\end{equation}

Where B$_\lambda$ is the Planck function and $\kappa_{ext}(\lambda) = \int_{r_{min}}^{\infty} n(r)\pi r^2 Q_{ext}(r,\lambda) \,dr$ is the total cloud opacity at $\lambda$ over all particle sizes (between 0.01 and 100~$\mu$m).
The total opacity is the sum of gas and cloud opacity $\kappa = \kappa_{R,g} +\kappa_{R,ext}$.
\bigbreak
For producing the pre-calculated cloud opacity tables, the wavelength dependent extinction coefficient, $Q_{ext}(r,\lambda)$ , single scattering albedo, $\omega(r,\lambda)$, and asymmetry factor, $g(r,\lambda)$, of each of our considered cloud species are calculated using the \textsc{MieX} Mie code of \citet{Wolf04} assuming spherical particles. 
A 2D grid of particle size and temperature dependent Rosseland weighted quantities is generated using these wavelengths dependent quantities, able to be interpolated inside the simulation to find the cloud properties for feedback into the RT scheme.
Real and imaginary optical constants for each of the cloud species are taken from \citet{Kitz18}.
The temperature-wavelength variation of the extinction and scattering coefficient and asymmetry factor for the six cloud composition considered are visible in the Appendix~\ref{App}.

\subsection{Cloud microphysical model}
\label{Sec:model3}

The transport of condensable gas particles is represented by two tracers one for the gas phase and one for the condensed phase. The condensation of the particles is performed as follows :

\begin{equation}
    \frac{dq_v}{dt} = -\frac{q_v - q_s}{\tau_c}\delta + \frac{min(q_s - q_v,q_c)}{\tau_c}(1 - \delta) + Q_{deep}
\end{equation}

\begin{equation}
    \frac{dq_c}{dt} = \frac{q_v - q_s}{\tau_c}\delta - \frac{min(q_s - q_v,q_c)}{\tau_c}(1 - \delta) + \frac{1}{\rho}\frac{\partial(\rho<q_c V_s>)}{\delta z}
\end{equation}

With q$_v$ the mass ratio of condensable vapor (kg~kg$^{−1}$), q$_c$ the mixing ratio of condensate (kg~kg$^{−1}$), q$_s$ is the local saturation vapor mixing ratio, V$_s$ is the settling speed of particles described in Eq~\ref{Eq9}, $\rho$ is the gas density, $\tau_c$ is the conversion timescale representing due to condensation or evaporation. The term Q$_{deep}$ = -(q$_v$ - q$_{deep}$)/$\tau_{deep}$ only applies to pressure greater than 50 bars, relaxing local vapor q$_v$ to the deep mixing ratio q$_{deep}$ over a characteristic timescale $\tau_{deep}$ set to 10$^3$~s. A sensitivity study of $\tau_{deep}$ was conducted in \cite{Tan19} over three order of magnitude for no substantial difference. The conversion timescale $\tau_c$ is set to 10~s \citep{Hell14}, short compared to cloud settling and radiative timescales. This conversion timescale will affect the depth of the convective layer, a larger value will allow vapor or condensate that should change phase to enrich the layer of the advecting plume is in. Values of 1~s and 100~s test were conducted, a small change of the cloud layer depth was observed. 
\bigbreak
This study is focused on brown dwarfs of type L and T clouds, in this temperature range we considered 4 representative cloud that could be affected by convective activity: enstatite, iron, perovskite, and corundum. Enstatite (MgSiO$_3$), is set to represent the silicate cloud, one of the most abundant condensates in L and T dwarfs. The total gas pressure P$_T$(bar) at which enstatite saturates as a function of temperature T and metallicity ([Fe/H]) is calculated as follows \citep{Viss10} :

\begin{equation}
    10^4/T = 6.26 - 0.35~log(P_T) - 0.70~[Fe/H]
\end{equation}

For the iron clouds, the equilibrium condensation temperature is set as: \citep{Viss10}

\begin{equation}
    10^4/T = 5.44 - 0.48~log(P_T) - 0.48~[Fe/H]
\end{equation}

To represent the Ti clouds, perovskite (CaTiO$_3$) is considered, defined as \citep{Wake17} :

\begin{equation}
    10^4/T = 5.125 - 0.277~log(P_T) - 0.554~[Fe/H]
\end{equation}

Corundum (Al$_2$O$_3$) is representing Al clouds \citep{Wake17} :

\begin{equation}
    10^4/T = 5.014 - 0.2179~log(P_T) + 2.264~10^{−3}~(log(P_T))^2 - 0.585~[Fe/H]
\end{equation}

For the chrome clouds, the equilibrium condensation temperature is set as \citep{Morl12} : 

\begin{equation}
    10^4/T = 6.576 - 0.486~log(P_T) - 0.486~[Fe/H]
\end{equation}

Alabandite (MnS) is considered \citep{Morl12}

\begin{equation}
    10^4/T = 7.447 - 0.42~log(P_T) - 0.84~[Fe/H]
\end{equation}

The saturation mixing ratio q$_s$ is then obtained with q$_s$ = P$_T$q$_{deep}$/p. The deep mixing ratio q$_{deep}$ relative to H and He$_2$ in solar abundance is set to 0.0026, 0.0012, 8.6~10$^{-6}$, 2.2~10$^{-4}$, 1.76~10$^{-5}$ and 2.1~10$^{-5}$ using respectively the molar fraction of Mg, Fe, Ca, Al, Cr and Mn for a solar metallicity \citep{Lodd03}.
\bigbreak
In the present study, only one cloud species is included in each simulation. While, in reality, multiple species of clouds may exist at the same location at a given, for simplicity, we only examine simulations including one species at a time. This may be a reasonable approximation in circumstances when one species dominates in opacity, and it allows us to identify the behavior of each species independently. Convection coupled to more comprehensive cloud models will be studied in future work, as some species may form in the same locations simultaneously, and the properties of most massive/thickest clouds would most likely dominate.
\bigbreak
The release of latent heat for the different cloud particles is not taken into account. In the temperature range of the atmosphere of this study and solar metallicity, the energy from the latent heat is several orders of magnitude lower than in the Earth moist tropics \citep{Zhan20}. With a GCM, \cite{Tan17} studied the effect of silicate latent heat and found that temperature perturbations from significant silicate condensation were on the order of only 1 K.
\bigbreak
The cloud particle number per dry airmass N$_c$ is assumed constant throughout the atmospheric column, the local cloud properties are then determined from this number. A lognormal particle size distribution is implemented in the models, extensively used to parameterize clouds in brown dwarfs atmospheres \citep{Acke01,Barm11,Morl12} : 

\begin{equation}
    n(r) = \frac{N_c}{\sqrt{2\pi}\sigma r}exp(-(\frac{ln(r/r_0)}{\sigma})^2)
\end{equation}

where r is the particle radius, n(r) = dN$_c$/dr is the number density distribution, r$_0$ is the reference radius and $\sigma$ is the nondimensionalized constant measuring the width of the distribution. The parameter $\sigma$ controlling the width of the size distribution is not constrained. In this study $\sigma$ is set to 1, \cite{Tan19} test the sensitivity to this parameter for values from 0.1 to 1.5, and found significant differences as the reference radius is changed. The reference radius r$_0$ is calculated by solving the q$_c$ = $\frac{4}{3} \pi \rho_c \int_{0}^{\infty} r^3n(r) \,dr$ = $\frac{4}{3} \pi \rho_c N_c r_0^3 exp(-\frac{9}{2}\sigma^2)$. A maximum radius is set to 100~$\mu$m, and a minimum radius is set to 0.01~$\mu$m \citep{Tsuj02}. 
\bigbreak
Cloud particles of radius $a$ are assumed to immediately reach their terminal fall speed given by \cite{Prup78}

\begin{equation}
    V_s = \frac{2 \beta a^2 g (\rho_p - \rho)}{9\eta}
    \label{Eq9}
\end{equation}

where $\eta$ is the viscosity of the gas, g is the gravitational acceleration of the planet, $\rho_p$ is the density of the particle and $\rho$ the density of the atmosphere. $\beta$ is the Cunningham slip that accounts for the gas kinetic effects, becoming relevant when the mean free path of the atmospheric molecules is bigger than the size of the falling particle. The expression of $\beta$ determined with experiments \citep{Li03} is 
\begin{equation}
    \beta = 1 + K_N(1.256 + 0.4e^{(-1.1/K_N)})
\end{equation}

Where K$_N$ is the Knudsen number, the ratio of 
the mean free path $\lambda$ to the size of the particle $a$, K$_N$=$\lambda$/a . For a perfect gas, the mean free path can be expressed as \citep{Chap70} 
 \begin{equation}
     \lambda = \frac{k_B T}{\sqrt{2}\pi d^2}\frac{1}{P}
 \end{equation}
 
for the viscosity $\eta$ of hydrogen the analytical formula from \cite{Rosn86} is used

\begin{equation}
    \eta = \frac{5}{16}\frac{\sqrt{\pi m k_B T}}{\pi d^2}\frac{(k_B T/\epsilon)^{0.16}}{1.22}
\end{equation}

with d the molecular diameter, m the molecular mass, and $\epsilon$ the depth of the Lennard-Jones potential well. m is set 2.827×10$^{-10}$~m and $\epsilon$ to 59.7k$_B$~K. This expression is valid for temperatures between 300~K and 3000~K and for pressures less than 100~bar \citep{Stie63}. 

\subsection{Simulation settings}

\begin{figure}[H]
 \center
 \includegraphics[width=14cm]{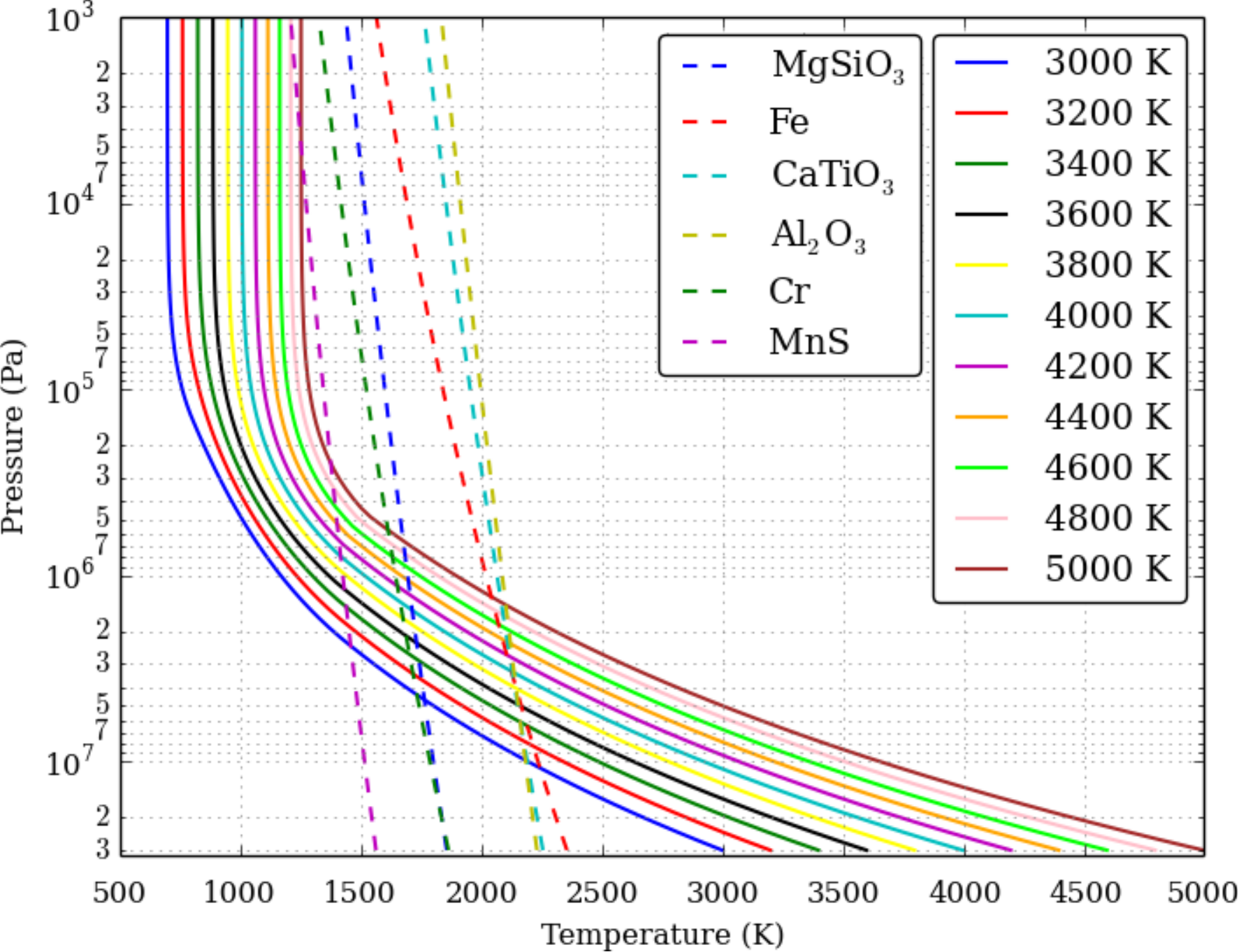}
 \caption{Initial vertical temperature profile (solid line) and condensation profile for the considered clouds (dotted line). The effective temperatures are for the clear sky cases.}
 \label{241}
\end{figure}

Fig~\ref{241} shows the initial temperature profiles from the \cite{Tan19} model and the condensation curve for the different cloud species considered. The horizontal resolution is 2~km over between 200 and 360 points, depending on the temperature case. The vertical domain extends from 3~10$^7$~Pa to 10$^4$~Pa for the lowest surface temperature case and 3~10$^3$~Pa for the highest. The time-step is 0.5~s. To avoid spurious reflection of upward propagating gravity waves on the model top, a Rayleigh damping layer where the energy is dissipated is applied over the 10 last levels with a damping coefficient of 1/3~10${-2}$~s$^{-1}$. The surface is set constant in space and time. We evaluate results after simulating 24~hours of time in cloudless cases, and 48~hours of time for cloudy cases. The vapor tracer is set at the start of the simulation at the deep mixing ratio on the first level, and the cloud tracer is set to zero. The vapor tracer is then advected vertically by the convection and condensates when it is possible. The different parameters are summarized in Table~\ref{T1}.

\begin{table}[!ht]
\center
\begin{tabular}{|l|cccc|}
\hline
Parameter & \multicolumn{4}{ |c| }{Value} \\
\hline
Gravity (m~s$^{-2}$) & \multicolumn{4}{ |c| }{1000} \\
Heat Capacity (J~K$^{-1}$) & \multicolumn{4}{ |c| }{13000} \\
Surface Pressure (Pa) & \multicolumn{4}{ |c| }{3~$\times$~10$^7$} \\
Mean molecular mass (g/mol) & \multicolumn{4}{ |c| }{2.23} \\
\hline
\hline
Clouds composition & \multicolumn{4}{ |l| }{\hspace{.2cm} MgSiO$_3$, \hspace{.3cm} Fe, \hspace{.5cm} CaTiO$_3$,\hspace{.3cm} Al$_2$O$_3$, \hspace{.5cm}Cr, \hspace{.6cm}MnS} \\
Deep mixing ratio (kg/kg) & \multicolumn{4}{ |c| }{2.6~10$^{-3}$, 1.2~10$^{-3}$, 8.6~10$^{-6}$, 2.2~10$^{-4}$, 1.76~10$^{-5}$ 2.1~10$^{-5}$} \\
\hline
\hline
Horizontal resolution (km) &\multicolumn{4}{ |c| }{2} \\
\hline
Surface temperature (K) & Effective temperature (K) & \multicolumn{3}{ |c| }{Domain} \\
\hline
\hspace{1.5cm} 3000 & 860 & \multicolumn{3}{ |c| }{200$\times$200$\times$200} \\
\hspace{1.5cm} 3200 & 940 & \multicolumn{3}{ |c| }{200$\times$200$\times$200} \\
\hspace{1.5cm} 3400 & 1010 & \multicolumn{3}{ |c| }{240$\times$240$\times$200} \\
\hspace{1.5cm} 3600 & 1090 & \multicolumn{3}{ |c| }{280$\times$280$\times$250} \\
\hspace{1.5cm} 3800 & 1160 & \multicolumn{3}{ |c| }{280$\times$280$\times$260} \\
\hspace{1.5cm} 4000 & 1230 & \multicolumn{3}{ |c| }{280$\times$280$\times$270} \\
\hspace{1.5cm} 4200 & 1290 & \multicolumn{3}{ |c| }{280$\times$280$\times$270} \\
\hspace{1.5cm} 4400 & 1350 & \multicolumn{3}{ |c| }{300$\times$300$\times$270} \\
\hspace{1.5cm} 4600 & 1400 & \multicolumn{3}{ |c| }{300$\times$300$\times$280} \\
\hspace{1.5cm} 4800 & 1450 & \multicolumn{3}{ |c| }{320$\times$320$\times$300} \\
\hspace{1.5cm} 5000 & 1500 & \multicolumn{3}{ |c| }{360$\times$360$\times$300} \\
\hline
\end{tabular}
\caption{Planetary and atmospheric parameters for the different simulations.}
\label{T1}
\end{table}

\section{Cloud free atmosphere}
\label{Sec:free}

\begin{figure}[H]
 \center
 \includegraphics[width=17cm]{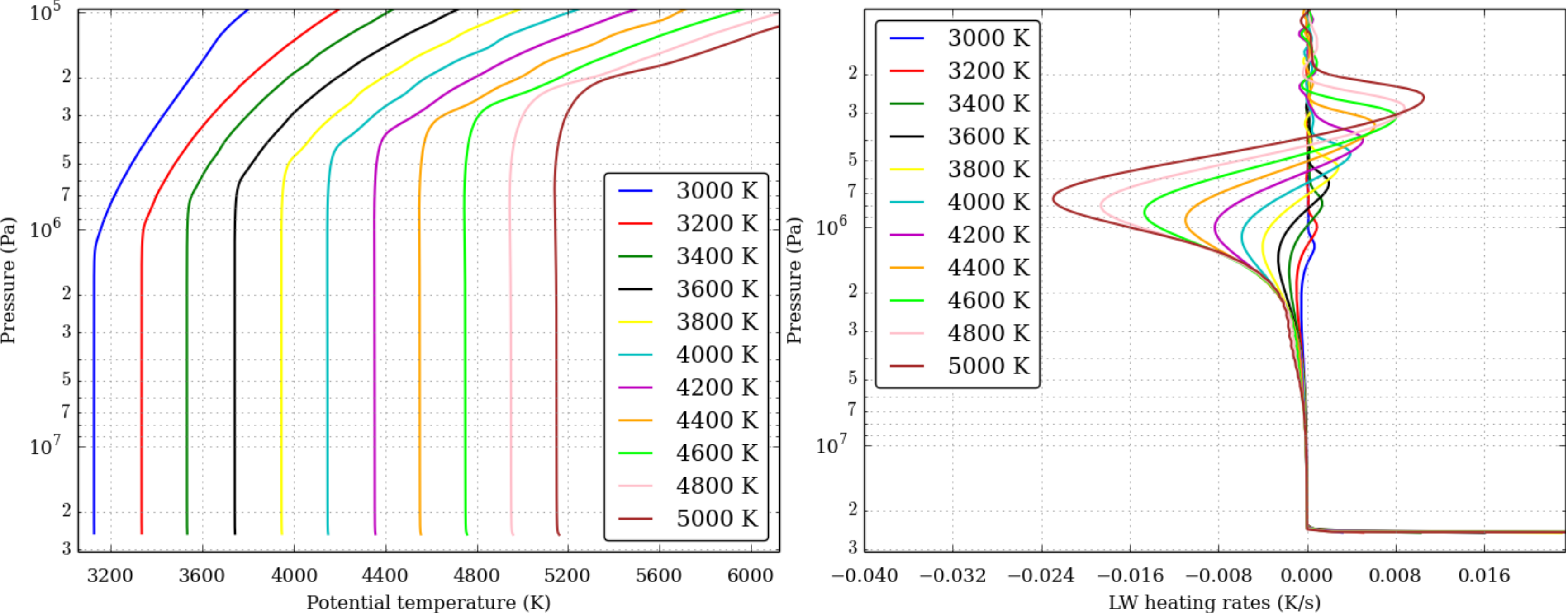}
 \caption{Domain averaged vertical profiles of the potential temperature (left) and IR heating rates (right) for the clear sky cases.}
 \label{31}
\end{figure}

Fig~\ref{31} displays the domain averaged vertical profiles of the potential temperature (left) and IR heating rates (right) for 10 different temperature cases, all cloud free for a solar metallicity value. The potential temperature is calculated using the heat capacity and the surface pressure in Table~\ref{T1} as reference pressure. The depth of the convective layer is visible in the constant value of the potential temperature, following a straightforward behaviour: The thickening of the convective layer is expected from the increase in opacity that comes with increasing temperature \citep{Pier10}. This is what is observed on the right panel on Fig~\ref{31}. Using a hydrostatic formulation, the depth of the convective layer in kilometers is comparable to the ones in \cite{Frey10}. The positive heating rates above 10$^6$~Pa are caused by the convective overshooting, which balance the excessive heating caused by convective penetration. This effect is important to determine the top of the convective zone, and is usually absent in 1D models.

\begin{figure}[H]
 \center
 \includegraphics[width=16.5cm]{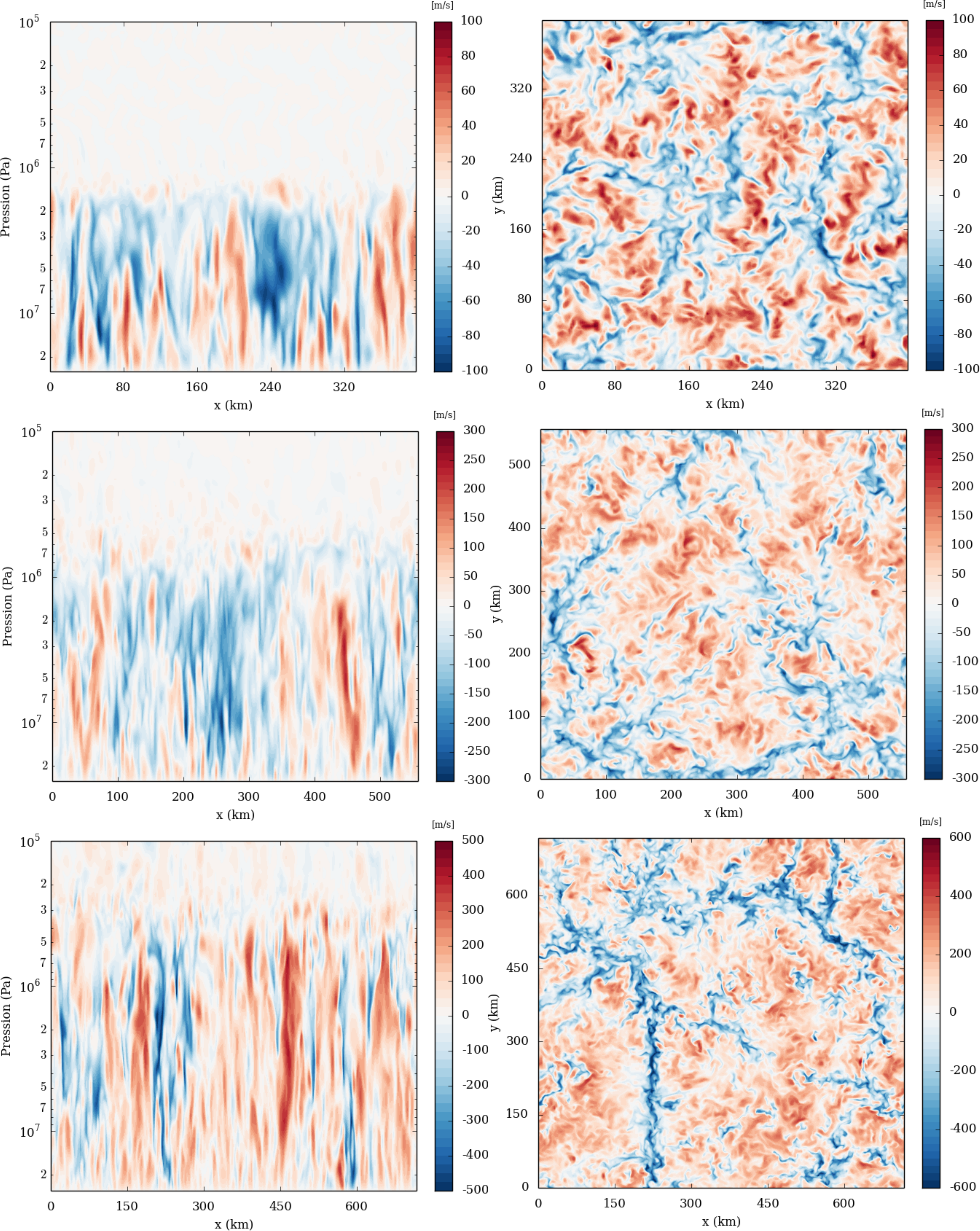}
 \caption{Snapshots of the vertical (left column) and horizontal (right column) cross-section of the vertical wind for the clear sky 3000~K case (top row), 4000~K case (middle row) and 5000~K case (bottom row). The horizontal cross-section is at 5.6~10$^{6}$~Pa for the 3000~K case, at 2.7~10$^{6}$~Pa for the 4000~K case and at 2~10$^{6}$~Pa for the 5000~K case.}
 \label{32}
\end{figure}

Fig~\ref{32} shows snapshots of the vertical (left column) and horizontal (right column) cross-section of the vertical wind for three clear-sky temperature cases: 3000~K (top row), 4000~K (middle row) and 5000~K (bottom row). The temperature trend on the convection depth has an impact on the convection depth, a hotter atmosphere will exhibit higher vertical wind, with a value 6 times higher between the two extreme temperature cases presented here. The convective layer organizes itself on the horizontal plane with polygonal cells with a diameter that will vary in temperature, from 80~km at 3000~K to 300~km at 5000~K. The values of the decimal logarithm value of the vertical velocity root-mean-square in cm~s$^{-1}$, 3.5 at 5.6~10$^{6}$~Pa for the 3000~K case, 3.9 at 2.7~10$^{6}$~Pa for the 4000~K, 4.2 at 2~10$^{6}$~Pa for the 5000~K case, are comparable to the \cite{Frey10} values with a similar temperature trend. This temperature trend is comparable to the scaling of vertical velocity with flux from mixing length theory \citep{Stev79}.

\begin{figure}[H]
 \center
 \includegraphics[width=12cm]{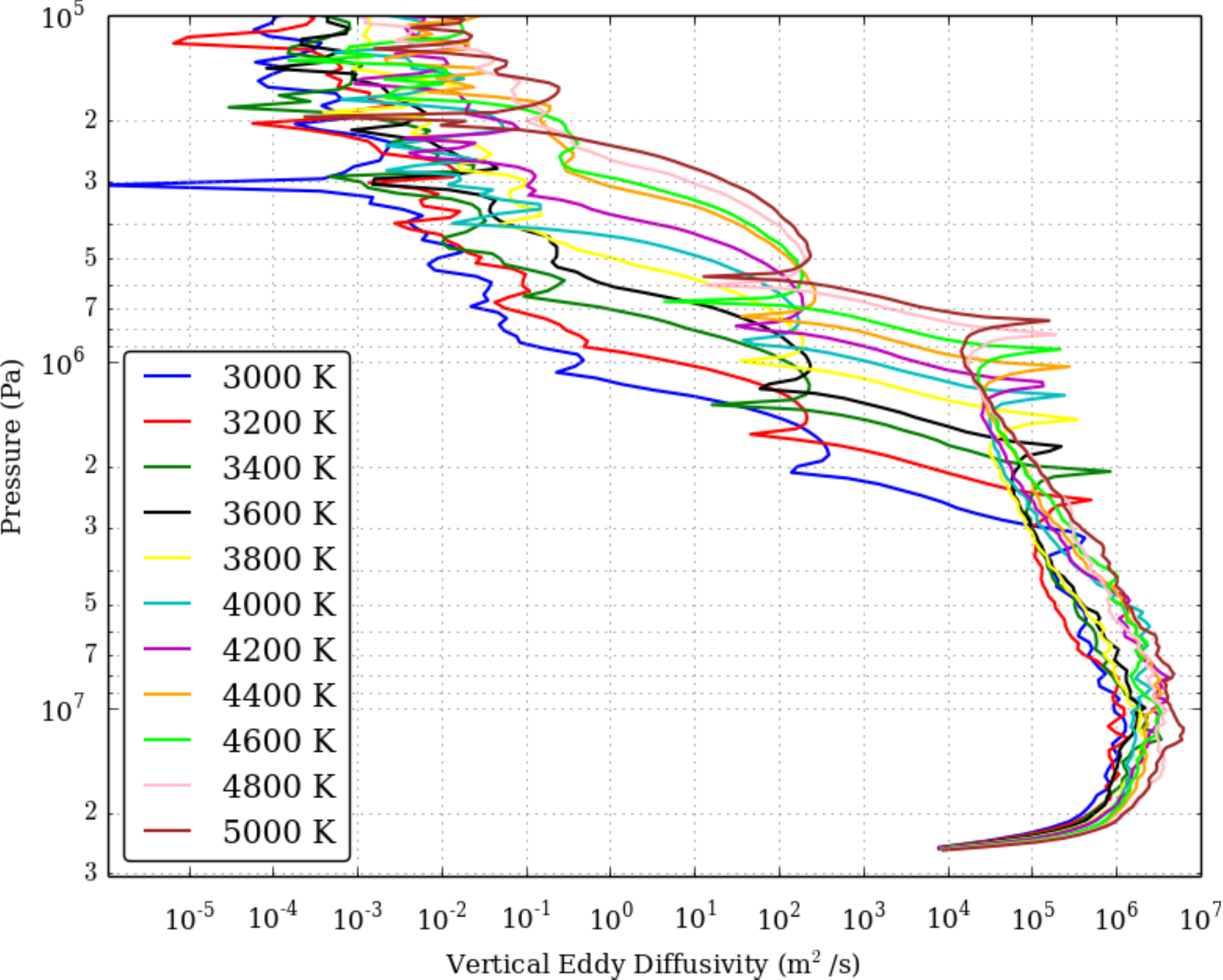}
 \caption{Domain averaged vertical profiles of the vertical eddy diffusivity (m$^2$/s) for the clear sky cases.}
 \label{33}
\end{figure}

Fig~\ref{33} displays the domain averaged vertical profiles of the vertical eddy diffusivity for the ten temperature cases calculated estimated as follows:
 \begin{equation}
     K_{zz} = -\frac{\overline{\theta'w'}}{\partial \overline{\theta}/\partial z}
 \end{equation}
With $\theta$ the potential temperature, w the vertical wind, primed quantities representing perturbations relative to the domain averaged values, and overlined quantities representing domain averaged values.
\bigbreak
The vertical eddy diffusivity exhibits the same temperature trend as the depth of convection, varying by a factor 3 between the two extreme temperature cases. The vertical eddy diffusivity has been estimated using tracer instead of potential temperature, and the values are of the same order of magnitude. The convective layer value is about 10$^6$ m$^2$/s, above there is the gravity waves participating into the diffusion but with several orders of magnitude lower that convection. This value of 10$^6$ m$^2$/s being higher than the value obtained by \cite{Frey10} with convection-resolving modelling. With mixing-length theory, the estimation of the vertical eddy diffusion is comparable with \cite{Frey10}. This value of the vertical eddy diffusivity in the convective layer is consistent with previous estimation \citep{Lewi10,Mose11}. In \cite{Tan19}, with a similar thermal structure and radiative transfer, the estimated vertical eddy diffusivity at 10$^6$~Pa is about the same order of magnitude. The scaling with temperature of the vertical eddy diffusivity is consistent with previous estimation \citep{Acke01}. Above the convective layer, the value of the vertical eddy diffusivity drops significantly and should not be used in GCMs or 1D simulations due to the absence of wind shear and general circulation.

\begin{figure}[H]
 \center
 \includegraphics[width=12cm]{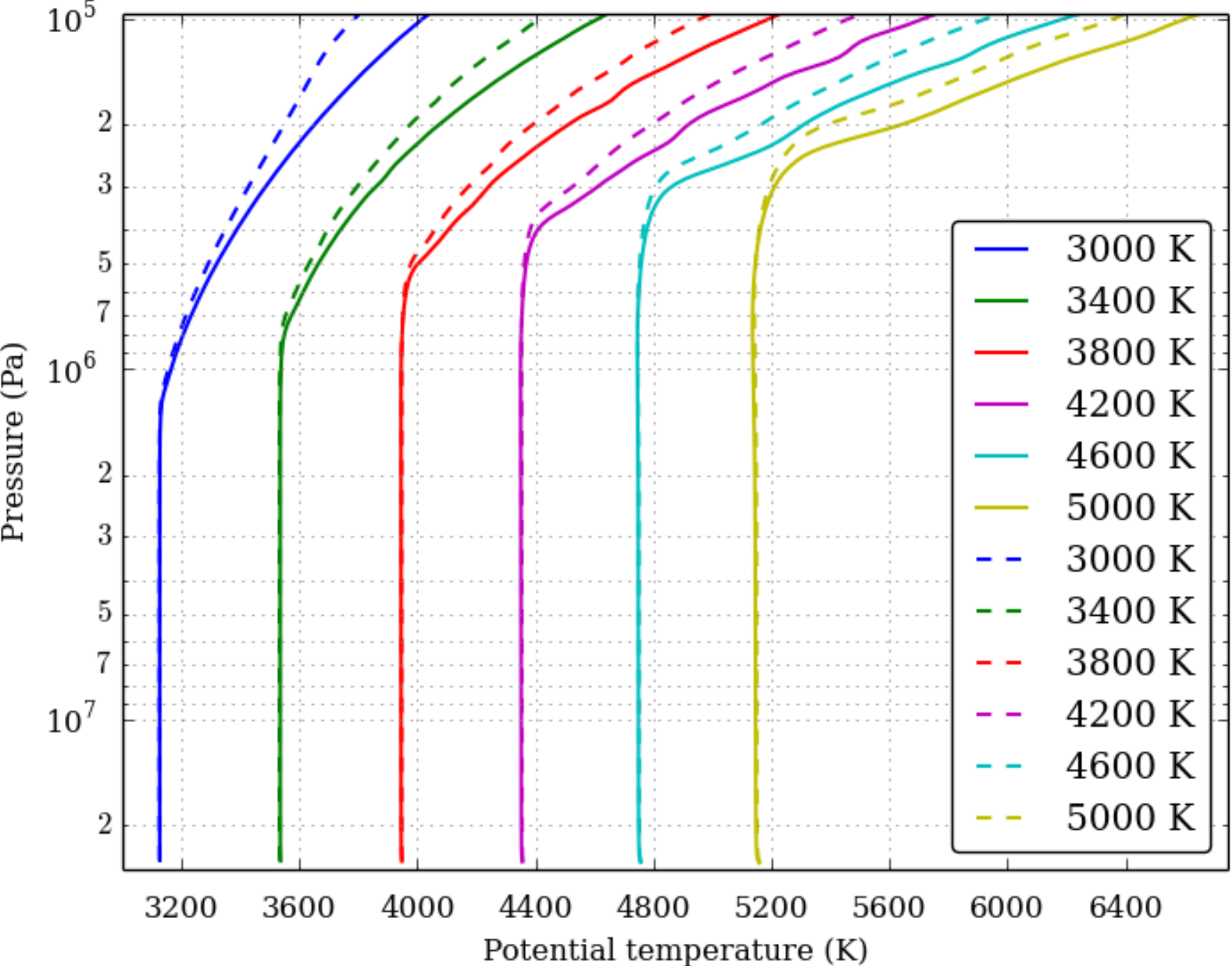}
 \caption{Domain averaged vertical profiles of the potential temperature for solar metallicity (solid lines) and 10 times the solar metallicity (dashed lines) for 6 clear-sky temperature cases.}
 \label{34}
\end{figure}

The \cite{Free14} gas opacity formulation takes into account the metallicity of the atmosphere, and simulations of clear-sky atmosphere with several metallicities has been performed. Fig~\ref{34} shows the impact of metallicity on the convective layer depth. The solid lines are for a solar metallicity value, similar to previous figures, and the dashed lines are for a metallicity equal to ten times the solar metallicity. The change of convection depth in barely noticeable at low temperature, and for the high temperature cases there is an increase with metallicity of the convective layer depth.

\section{Impact of Cloud}
\label{Sec:cloud}

\begin{figure}[H]
 \center
 \includegraphics[width=17cm]{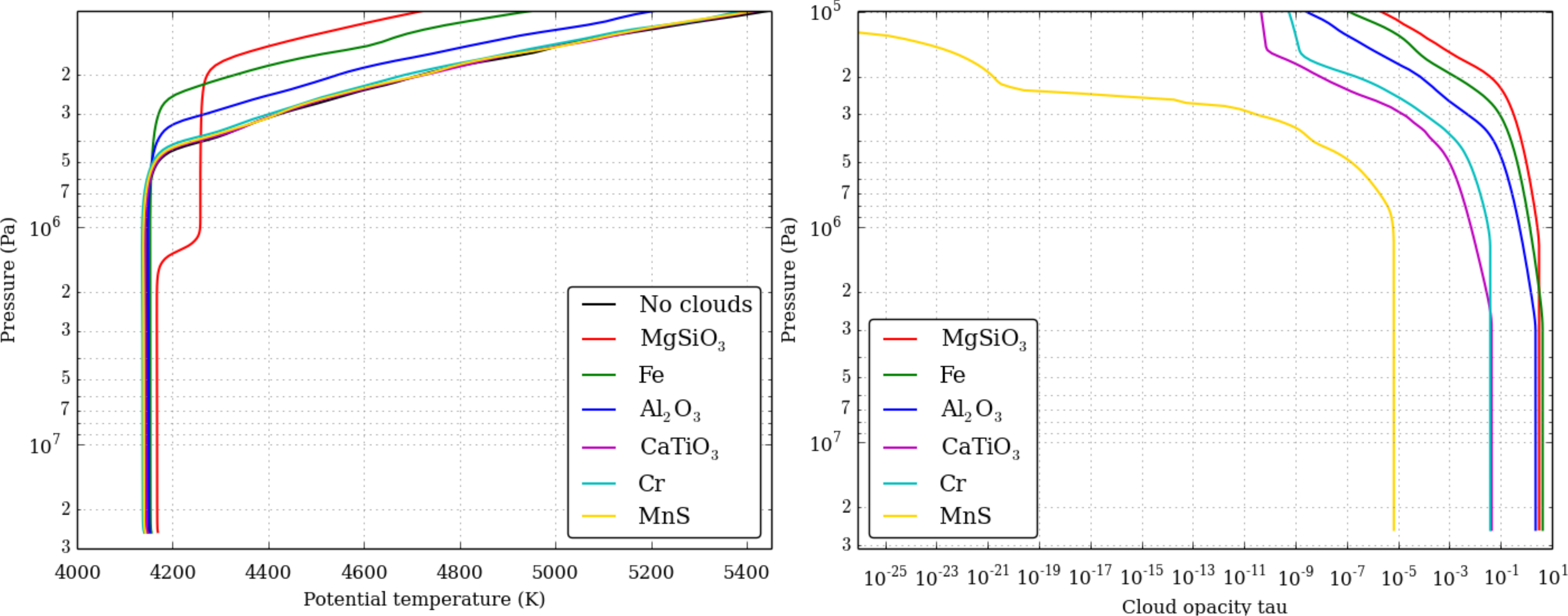}
 \caption{Domain averaged vertical profiles of the potential temperature (left) and cloud opacity (right) for the 6 cloud composition considered with a cloud particle number density of 10$^8$~kg$^{-1}$ for the 4000~K case.}
 \label{401}
\end{figure}

Fig~\ref{401} displays the Domain averaged vertical profiles of the potential temperature (left) and cloud opacity (right) for the 6 cloud compositions considered in this study for 10$^8$~kg$^{-1}$ cloud particle number per dry airmass and one temperature case. Three categories can be discerned, the clouds with opacity less than 1 and no impact on the convective layer composed of CaTiO$_3$, Cr and MnS, the clouds with opacity greater than 1 and moderate impact composed of Fe and Al$_2$2O$_3$, and the clouds with opacity greater than 1 and high impact composed of MgSiO$_3$. A small cloud particle number per dry airmass will increase the horizontal variability and can in some cases engender cloud holes. The $\tau_c$ conversion timescale will affect the cloud layer vertical thickness. With a larger timescale value, cloud particles will be able to mix to deeper layers before they evaporate.

\subsection{MgSiO$_3$ clouds}
\label{Sec:cloud1}

\begin{figure}[H]
 \center
 \includegraphics[width=17cm]{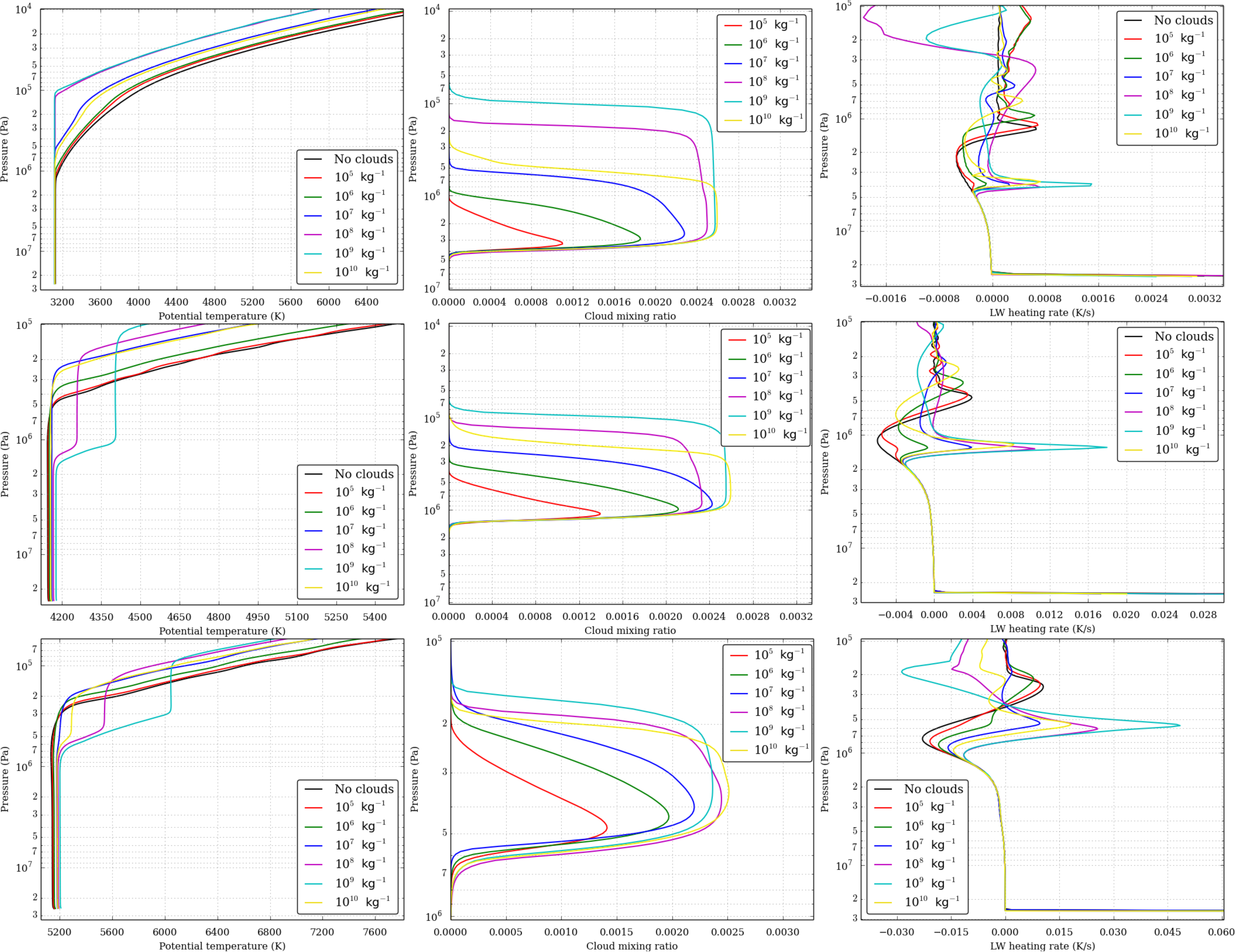}
 \caption{Domain averaged vertical profiles of the potential temperature (left column), cloud mixing ratio (middle column) IR heating rates (right column) for the 3000~K (top row), 4000~K (middle row) and 5000~K (bottom row) case with MgSiO$_3$ clouds and particle number per dry airmass between 10$^5$ and 10$^{10}$~kg$^{-1}$.}
 \label{411}
\end{figure}

Fig~\ref{411} shows on the left column the domain averaged vertical profiles of the potential temperature, cloud mixing ration and IR heating rates for three temperatures cases with MgSiO$_3$ clouds and six different particle number per dry airmass values. The associated domain averaged vertical profiles of reference particle radius, scattering albedo and opacity are visible in Appendix~\ref{App2}, Fig~\ref{A21}.

\begin{figure}[H]
 \center
 \includegraphics[height=23cm]{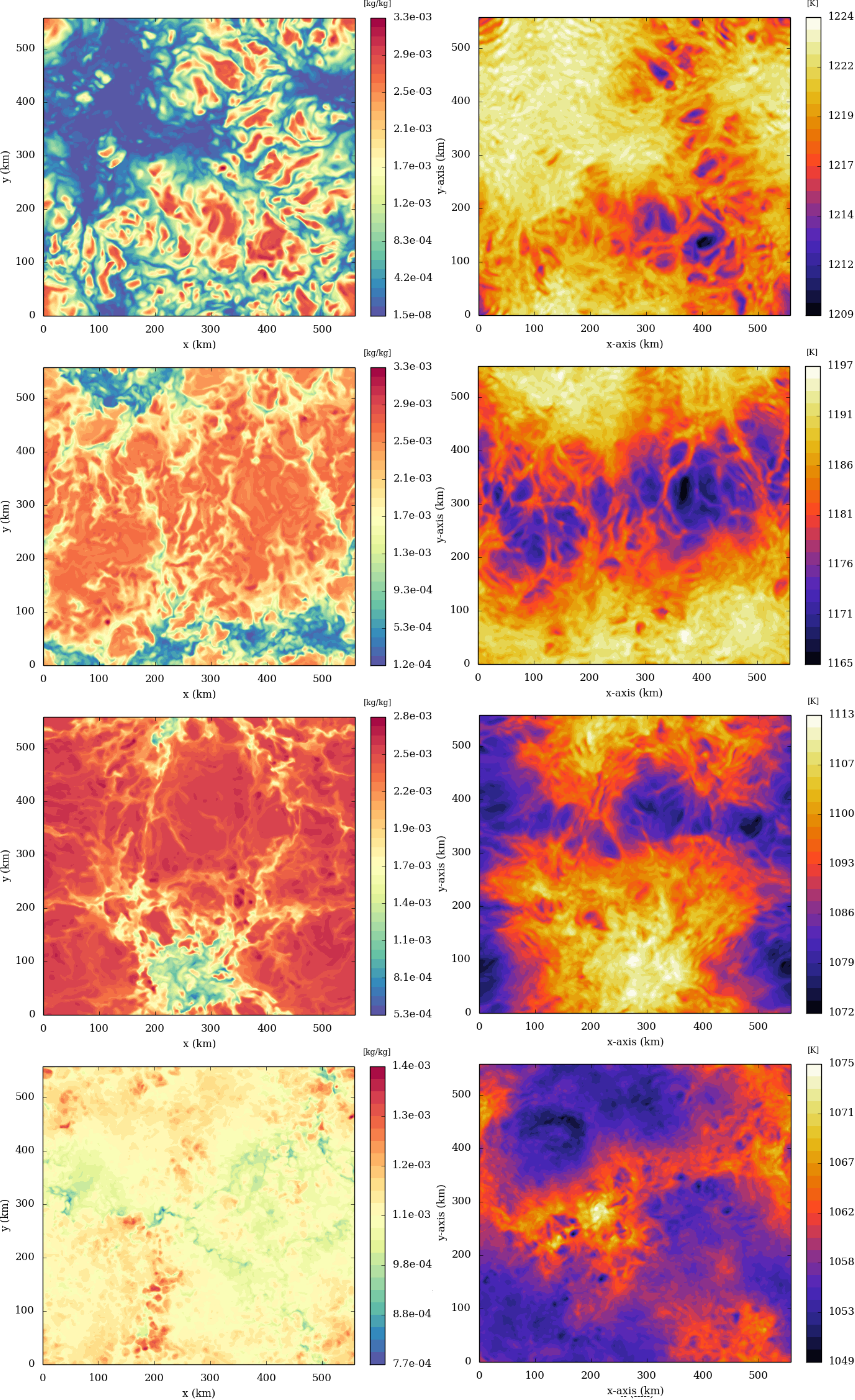}
 \caption{Snapshots of horizontal cross-section at 5.4~10$^5$~Pa of the cloud mixing ratio (left column) and the brightness temperature (right column) for the 4000~K case with MgSiO$_3$ clouds and cloud particle number per dry airmass from top to bottom of 10$^5$~kg$^{-1}$, 10$^6$~kg$^{-1}$, 10$^7$~kg$^{-1}$ and 10$^8$~kg$^{-1}$.}
 \label{412}
\end{figure}

The depth of the convective layer is strongly affected by the presence of MgSiO$_3$ clouds, with a strong increase of the top boundary of the convection. The presence of cloud is visible in the IR heating rate with a strong clouds-induced heating at 4~10$^6$~Pa for the 3000~K case, 1.5~10$^6$~Pa for the 4000~K case and at 6~10$^5$~Pa. This increase is different for the five different particle number per dry airmass values considered. For the low values, 10$^5$ and 10$^6$~kg$^{-1}$, the cloud particle reference radius is larger, above 10~$\mu$m, resulting in small opacities and small scattering albedo and therefore to a limited impact on the atmosphere. The IR heating vertical profiles are hardly changed for the three temperature cases shown here. The two cases of 10$^7$ and 1e$^{10}$~kg$^{-1}$ particle number per dry airmass have a similar impact on the convection depth, despite a difference in radii. The 1e$^7$~kg$^{-1}$ case leads to a reference radius around 5~$\mu$m, resulting in cloud particle with small opacities but high scattering albedo. Whereas The 1e$^{10}$ case leads to a small reference radius around 0.5~$\mu$m, resulting in small scattering albedo but higher opacities. The overall increase on the convection depth is similar. Finally, the 10$^7$ and 10e$^{10}$~kg$^{-1}$ particle number per dry airmass cases have strong impacts. With reference radius 1 and 2.5~$\mu$m, the clouds have high opacities, greater than 1 for the three temperature cases, and high scattering albedo, up to 0.7, resulting in strong heating. This heating leads in some cases to a detached convective layer, a secondary convective layer on top of the deep convective layer induced by the deep atmosphere with a different potential temperature than the deep atmosphere. This impact on the depth of the convective layer will engender a modification of the thermal structure above this layer and therefore a change in the effective temperature. For the 3000~K case, the effective temperature is equal to 860~K and drops to an average of 840~K, 830~K, 780~K, 690~K, 680~K and 790~K for respectively a cloud particle number per dry airmass of 10$^5$~kg$^{-1}$, 10$^6$~kg$^{-1}$, 10$^7$~kg$^{-1}$, 10$^8$~kg$^{-1}$, 10$^9$~kg$^{-1}$ and 10$^{10}$~kg$^{-1}$. For the 4000~K case, the effective temperature is equal to 1230~K and drops to an average of an average of 1170~K, 1190~K, 1080~K, 1050~K, 990~K and 1090~K for respectively a cloud particle number per dry airmass of 10$^5$~kg$^{-1}$, 10$^6$~kg$^{-1}$, 10$^7$~kg$^{-1}$, 10$^8$~kg$^{-1}$, 10$^9$~kg$^{-1}$ and 10$^{10}$~kg$^{-1}$. For the 5000~K case, the effective temperature is equal to 1500~K and drops to an average of 1475~K, 1440~K, 1390~K, 1340~K, 1330~K and 1370~K for respectively a cloud particle number per dry airmass of 10$^5$~kg$^{-1}$, 10$^6$~kg$^{-1}$, 10$^7$~kg$^{-1}$, 10$^8$~kg$^{-1}$, 10$^9$~kg$^{-1}$ and 10$^{10}$~kg$^{-1}$. As expected, the cases in which there is a detached convective layer are the one with the largest temperature decrease.
\bigbreak
Fig~\ref{412} shows snapshots of the cloud mixing ratio for the 4000~K case with MgSiO$_3$ clouds for four different cloud particle number per dry airmass and associated brightness temperature. The effect of the cloud particle number per dry airmass on the cloud is visible, the cloud tends to be more non-homogeneous and cloud holes are visible in the 10$^5$~kg$^{-1}$ case (top row), are not present for the 10$^8$~kg$^{-1}$ case. The 10$^9$~kg$^{-1}$ and 10$^{10}$~kg$^{-1}$ cases are not shown for clarity, but the homogenisation of the cloud is stronger than for the 10$^8$~kg$^{-1}$ case. With a small cloud particle number per dry airmass, the particle diameter is bigger (Fig~\ref{A21}) and therefore the terminal settling velocity will be higher by several orders of magnitude. For the low cloud particle number per dry airmass cases, it becomes significant relative to the convective vertical velocity, creating a competition between the two, resulting in cloud holes. When the bulk setting velocity is comparable to the flow vertical velocity, the settling timescale is comparable to the vertical advection timescale. Thick clouds are thus sustained only in regions of updrafts, whereas in downdraft regions clouds settle out quickly, generating significant cloud patchiness. In simulations with bulk setting velocity much smaller, for a high cloud particle number per dry airmass (below 10$^7$~kg$^{-1}$) than the flow vertical velocity, horizontal mixing effectively homogenize the cloud structure in the domain. The effect of the complete cloud coverage is also visible in the brightness temperature, with high cloud particle number values, the outgoing radiative flux is lower.

\begin{figure}[H]
 \center
 \includegraphics[width=17cm]{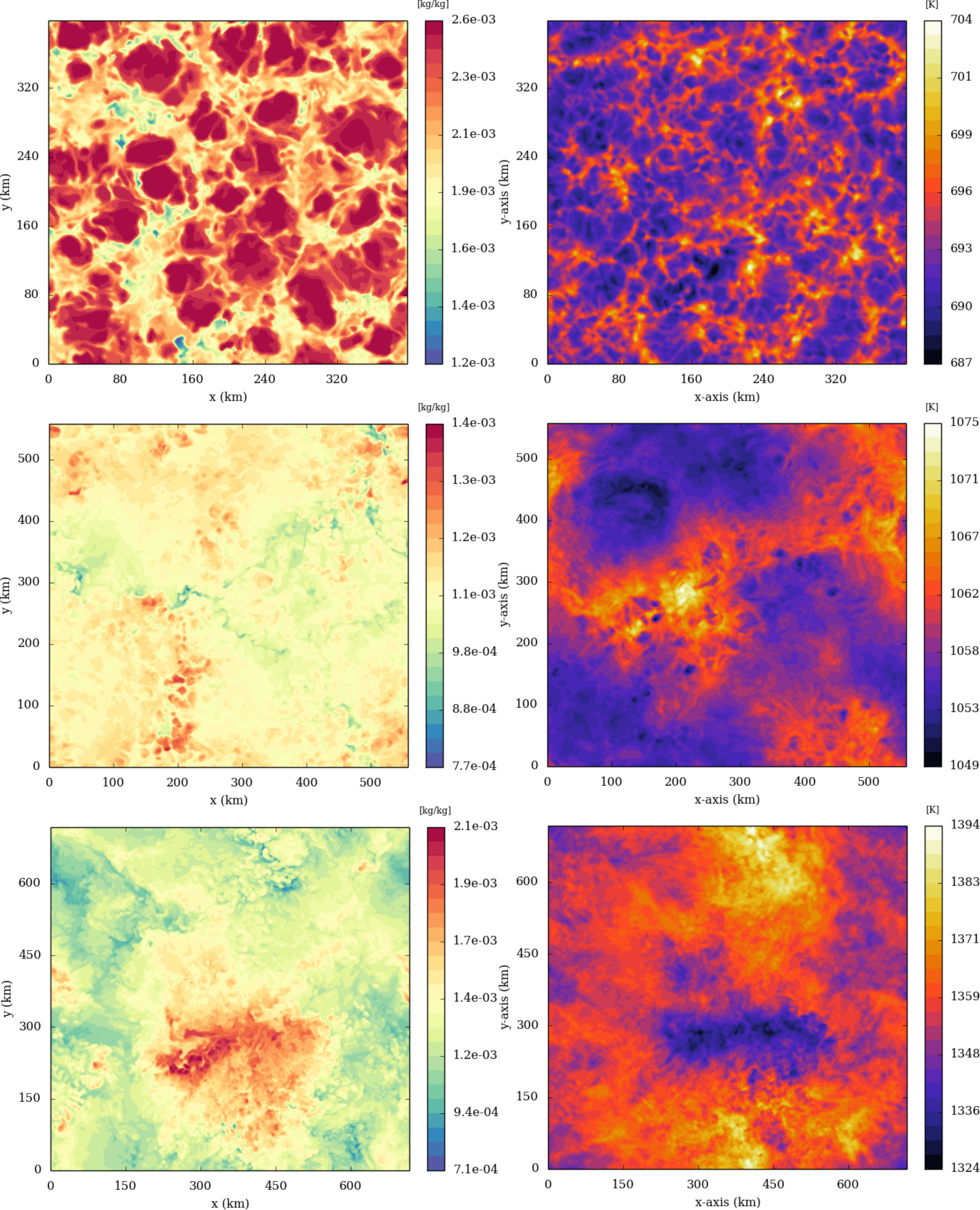}
 \caption{Snapshots of the horizontal cross-section of the cloud mixing ratio (left column) and the brightness temperature (right column) for the 3000~K case (top row), the 4000~K case (middle row) and the 5000~K case (bottom row) with MgSiO$_3$ clouds and cloud particle number per dry airmass of 10$^8$~kg$^{-1}$.}
 \label{413}
\end{figure}

Fig~\ref{413} shows snapshots of the cloud mixing ratio and brightness temperature for the three temperature cases with MgSiO$_3$ clouds for a 10$^8$~kg$^{-1}$ cloud particle number per dry airmass. In the three cases, the mixing ratio of the cloud particles are pretty uniform with noticeable cellular features. The cloud pattern changes with increasing temperature, showing some tendency to convective aggregation at high temperature. For the 3000~K case, the cloud holes are small, cells of few tens of kilometers, whereas for the 5000~K, the cloud holes are dominated by a 400~km diameter structure. This is due to the convective structure, as shown in Fig~\ref{32} there is an increase of the convective cell diameter with increasing temperature due to the thickening of the convective layer. At low temperature, the downdraft region is narrow whereas with the increase of temperature, the downdraft region where several updrafts meet can be quite large, leading to a cloud hole with the settling velocity. At low cloud particle number and high temperature, there can be large cloud holes of several hundreds of kilometers in diameter.

\subsubsection{Detached convective layer}

\begin{figure}[H]
 \center
 \includegraphics[width=17cm]{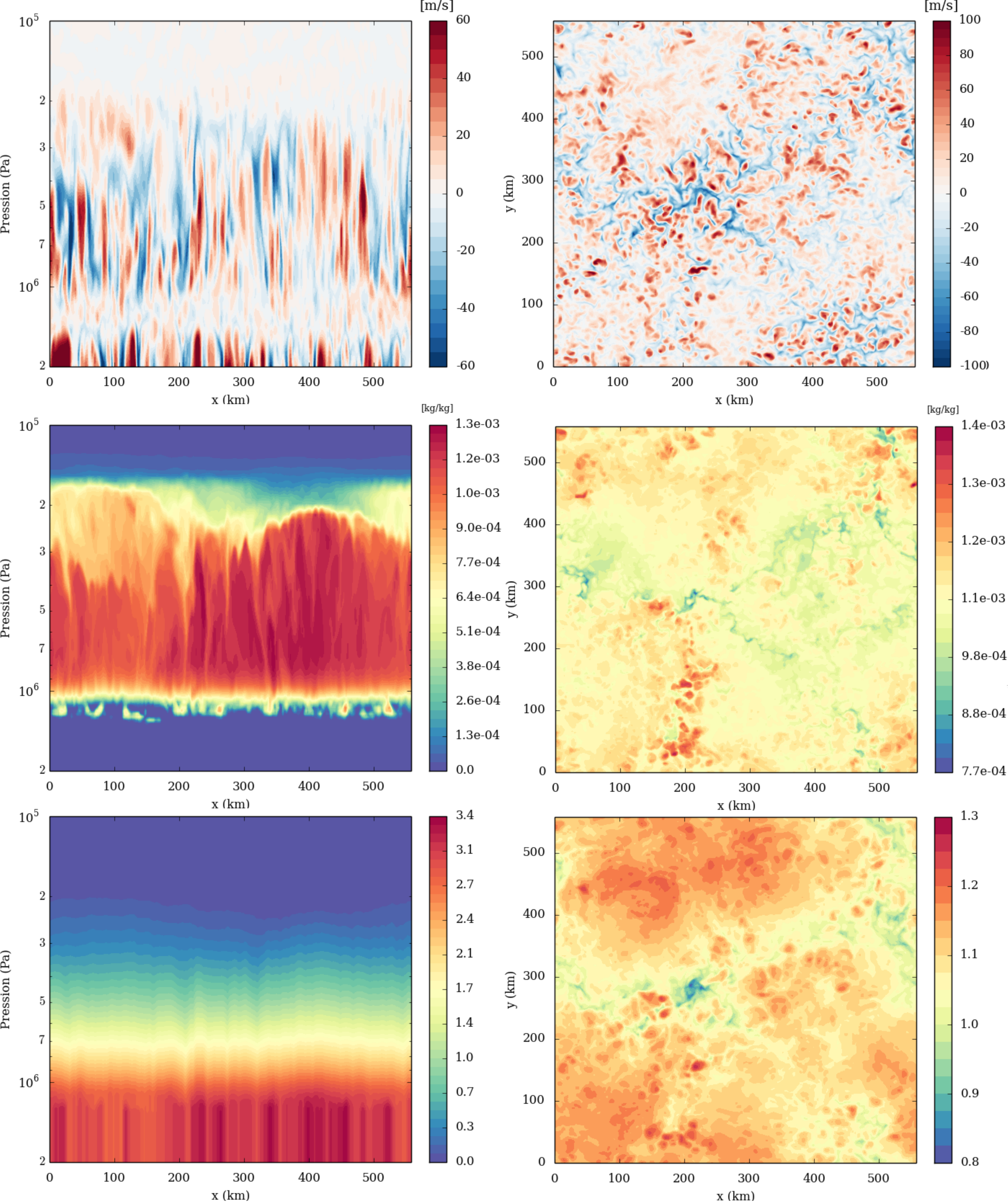}
 \caption{Snapshots of the vertical cross-section (left column) and horizontal cross-section at 5.4~10$^5$~Pa (right column) of the vertical wind (top row), cloud mixing ratio (middle row) and cloud opacity (bottom row) for the 4000~K case with MgSiO$_3$ clouds and cloud particle number per dry airmass of 10$^8$~kg$^{-1}$.}
 \label{414}
\end{figure}

These detached convective layers are triggered by an intense IR heating induced by the clouds that destabilize the atmosphere and engender a separate convective layer visible on the left column of Fig~\ref{411} by an area above 10$^6$~Pa with distinctive constant potential temperature. This strong cloud greenhouse effect produces a stratified layer below the cloud base, which splits the convection zone into two. The root cause behind this effect is that cloud opacity decreases rapidly with increasing pressure near the cloud base, and this configuration favors a stratified layer. In the 3000~K case, there is a detached convective layer for a particle number density of 10$^8$~kg$^{-1}$ that is not discernible in the figure. In the 4000~K case, the particle number density values of 10$^8$ and 10$^9$~kg$^{-1}$ generate a detached convective layer; whereas for the 5000~K, a detached convective layer is present for the 10$^8$, 10$^9$ and 10$^{10}$~kg$^{-1}$ particle number density cases. Fig~\ref{414} shows instantaneous snapshots of the detached convective layer for the 4000~K case and a particle number density of 10$^8$~kg$^{-1}$. This cloud greenhouse effect on the detached zones has been seen in many previous 1D models with clouds \citep{Burr06,Tan19}.

\begin{figure}[H]
 \center
 \includegraphics[width=17cm]{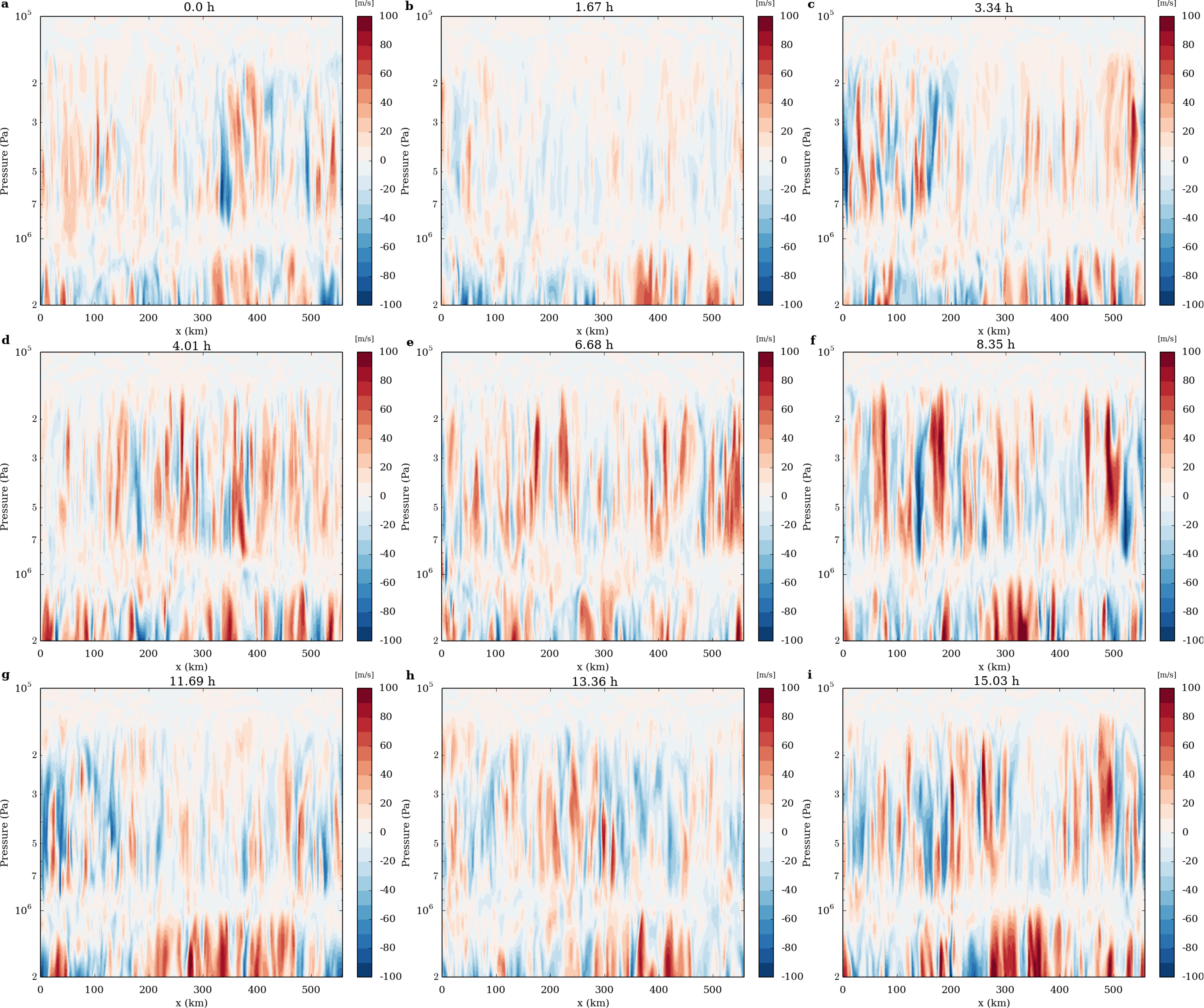}
 \caption{Time-series of vertical cross-sections of the vertical wind for the 4000~K case with MgSiO$_3$ clouds and cloud particle number per dry airmass of 10$^8$~kg$^{-1}$. The frequency between the panels is 1.67~h.}
 \label{415}
\end{figure}

The detached convective layer is visible between 10$^6$ and 2~10$^5$~Pa, below the deep convective layer is visible in the vertical cross-section of the vertical wind. The associated horizontal cross-section shows the organization of the convection. Contrary to the deep convective layer organized in closed-polygonal cells (see Fig~\ref{32}), the detached convective layer is organized with weak large polygonal updrafts (see Fig~\ref{414}-top right panel at x~=~200~km and y~=~450~km) surrounded by strong small updrafts with positions varying in time (see Fig~\ref{414}-top right panel at x~=~200~km and y~=~300~km). The heating due to clouds scattering at the base of this detached layer is strong enough to maintain the convective layer detached but not enough to sustain a polygonal cell organization, resulting in a puffiness of convective updraft and clouds. The vertical eddy diffusion is lower (between 10$^3$ and 10$^4$~m$^2$/s) for the cases showed in Fig~\ref{414}, lower than in the deep convective layer due to the non-homogeneity of the convective plume strength. The position of the strong updraft region evolves in time. The height and thickness of the detached convective layer depend on the heating induced by the cloud, the stronger it is, the higher and thicker the convective layer will be. The metallicity also has an effect on the detached convective layer.
\bigbreak
The temporal variability of this detached convective layer with a time series of vertical cross-section of the vertical wind is shown in Fig~\ref{415} over 15~h. The deep convective activity is visible below 10$^6$~Pa. Comparing the two convective layers, there is a strong variability in the detached convective layer. Fig~\ref{415}-b and Fig~\ref{415}-f represents two extreme cases, where for the first one there is a tenuous convective activity with low vertical wind and on the contrary, for the second one there is a substantial convective activity with vertical wind similar in amplitude to the deep convection. In-between those two, there are transient steps with regions without barely any convective activity and regions with significant convection. The frequency of occurrence of those convective plumes is roughly about 7-8~h. This frequency is of the same order of magnitude as in \cite{Tan19}, over several hours, however the domain-mean variability on the potential temperature is lower. The impact on the clouds is also weaker, the temporal variation of the cloud abundance is inferior by one order of magnitude. This might be due to the vertical mixing inside this detached layer in the \cite{Tan19} 1D model, where it is stronger by at least a factor 10.

\subsection{Fe and Al$_2$O$_3$ clouds}

\begin{figure}[H]
 \center
 \includegraphics[width=17cm]{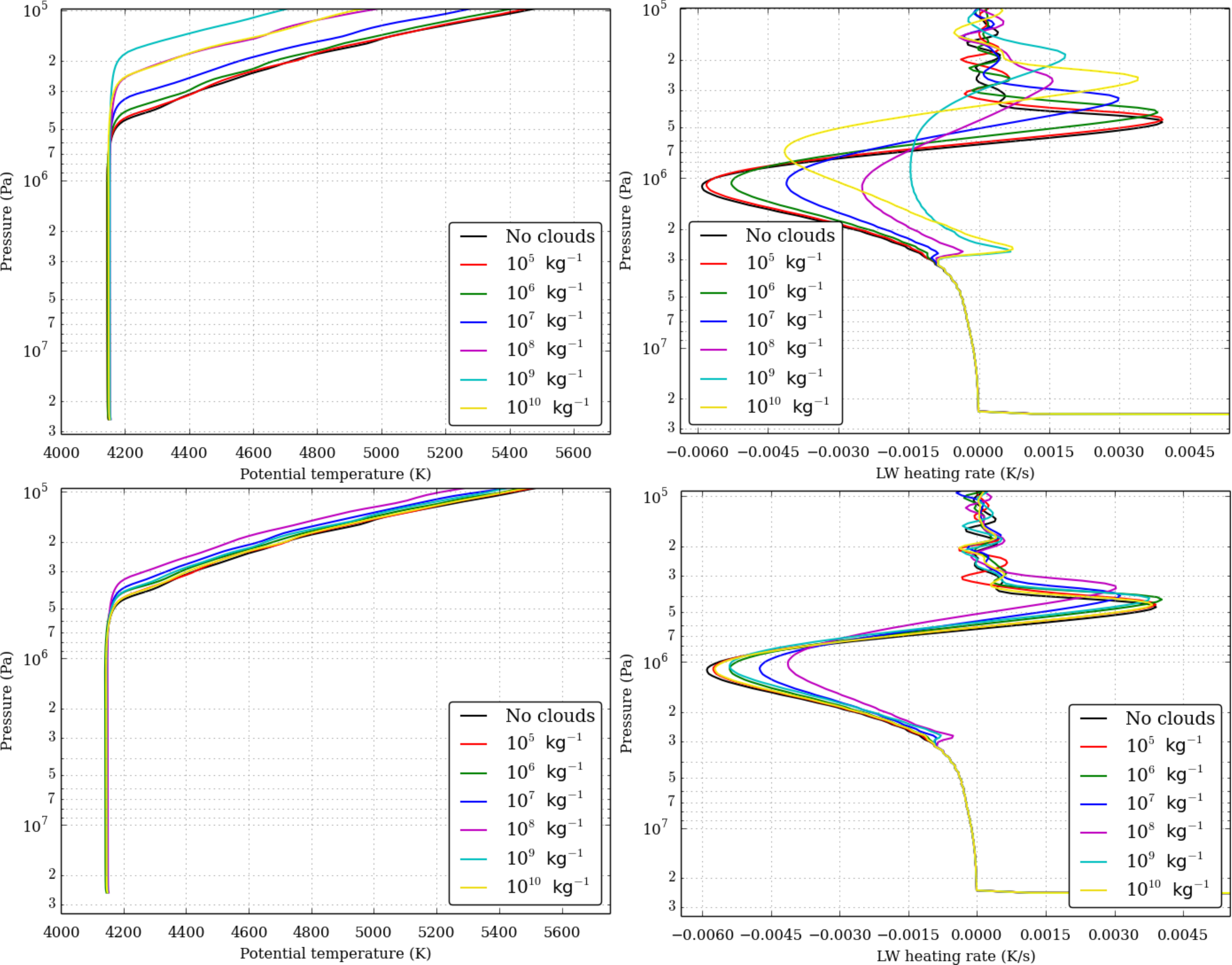}
 \caption{Domain averaged vertical profiles of the potential temperature (left) and IR heating rates (right) for the 4000~K case with Fe (top row) and Al$_2$O$_3$ (bottom row) clouds and particle number per dry airmass between 10$^5$ and 10$^{10}$~kg$^{-1}$.}
 \label{421}
\end{figure}

In Fig~\ref{401}, the impact of Fe and Al2O$_3$ on the convective layer depth is lower than that of MgSiO$_3$ clouds, but is still noticeable. Fig~\ref{421} shows on the left column the domain averaged vertical profiles of the potential temperature and IR heating rates for the 4000~K temperature case with Fe (top) and Al$_2$O$_3$ (bottom) clouds and 6 different particle number per dry airmass values. The associated domain averaged vertical profiles of reference particle radius, scattering albedo and opacity are visible in Fig~\ref{A31}. As for MgSiO$_3$, the particle number plays a role. The bigger particle, 10$^5$ and 10$^6$~kg$^{-1}$ cloud particle number density cases, barely affect the convective layer. The four other cases have different effects depending on the cloud particle species. For Fe clouds, the 10$^9$~kg$^{-1}$ case will engender the strongest increase of the convective layer depth due to the higher scattering albedo and opacity. However, for Al$_2$O$_3$, the value of the particle number per dry airmass of which the convective depth is maximal is 10$^8$~kg$^{-1}$. The reference radius of these two cases is about 1~$\mu$m.

\subsection{CaTiO$_3$, Cr and MnS clouds}

\begin{figure}[H]
 \center
 \includegraphics[width=17cm]{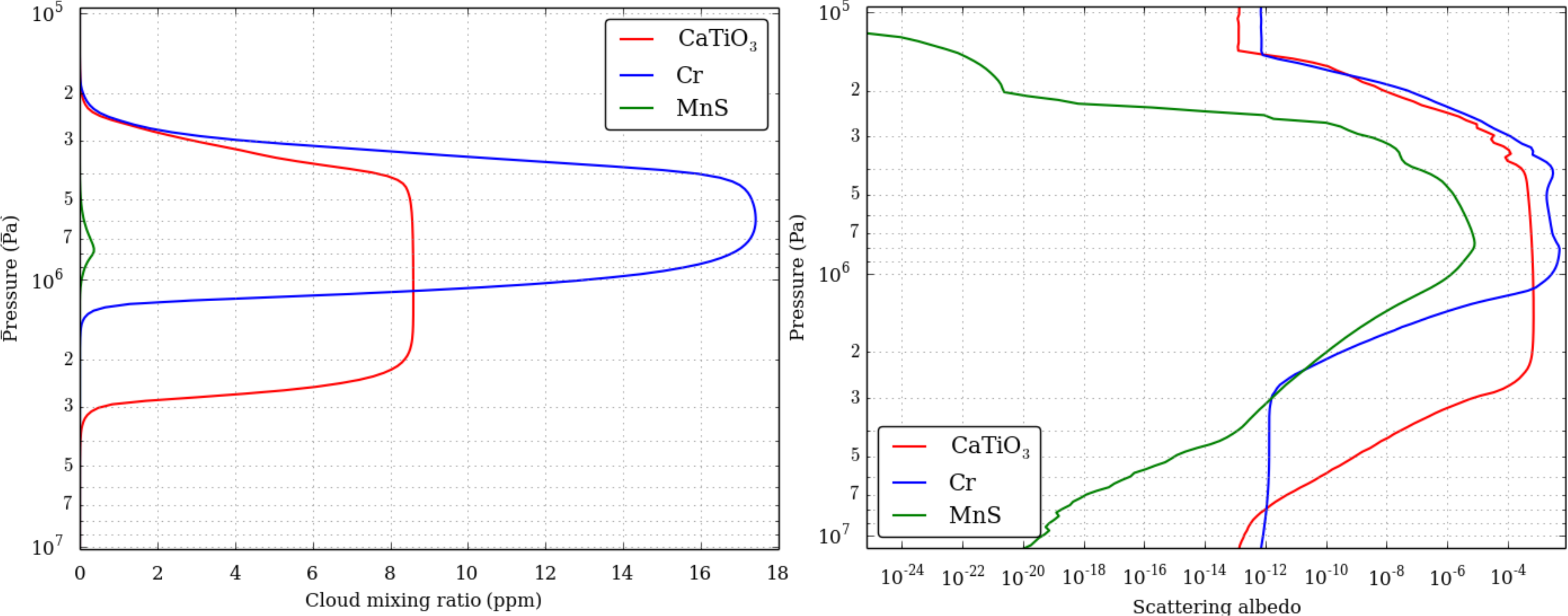}
 \caption{Domain averaged vertical profiles of the cloud particles mixing ratio (left) and scattering albedo (right) for the 4000~K case with CaTiO$_3$, Cr and MnS clouds and particle number per dry airmass of 10$^8$~kg$^{-1}$.}
 \label{431}
\end{figure}

In Fig~\ref{401}, the CaTiO$_3$, Cr and MnS clouds showed hardly any impact on the convective layer depth. Fig~\ref{431} shows the domain averaged cloud mixing ratio and scattering albedo vertical profiles for the three cloud species. The very small impact of CaTiO$_3$, Cr and MnS clouds are due to two factors, the mixing ratio and the vertical thickness of the cloud layer. The deep mixing ratio value for Ca, Cr and Mn are 8.6~10$^{−6}$, 1.76~10${−5}$ and 2.1~10${−5}$, between a factor 10 and 100 lower than Al, where Al$_2$O$_3$ has small impact. CaTiO$_3$ clouds have a similar depth than Al$_2$O$_3$ clouds, however Cr and especially MnS will have a cloud layer significantly thinner. For MnS, the cloud layer will be present only at the very top of the convective layer and the mixing will be therefore less efficient compared to other clouds composition considered here. These two factors will lead to low opacity and low scattering albedo (meaning an increase of the outgoing infrared radiation compared to others clouds), and an overall limited impact on the convection depth. 

\subsection{Impact of cloud metallicity}

\begin{figure}[H]
 \center
 \includegraphics[width=17cm]{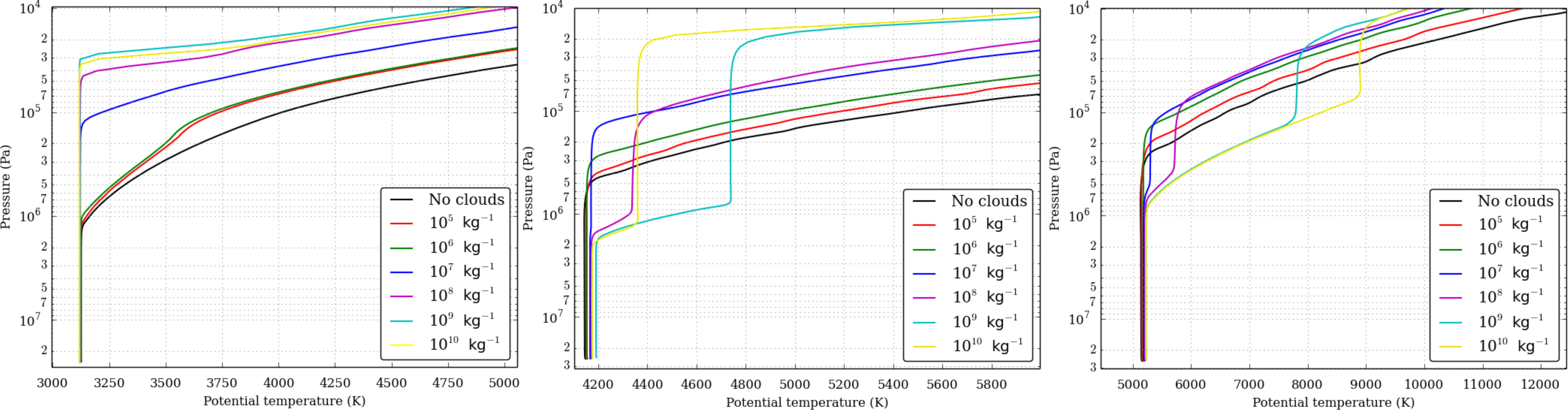}
 \caption{Domain averaged vertical profiles of the potential temperature for the 3000~K (left), 4000~K (middle) and 5000~K (right) case with MgSiO$_3$ clouds and particle number per dry airmass between 10$^5$ and 10$^{10}$~kg$^{-1}$ with solar metallicity multiplied by 10.}
 \label{416}
\end{figure}

Fig~\ref{34} showed the impact of the metallicity of the background atmosphere, and in this section is presented the impact of metallicity of the deep mixing ratio q$_{deep}$ discussed in Section~\ref{Sec:model3}. Fig~\ref{416} shows the potential temperature for three temperature cases considered with MgSiO$_3$ and ten times the solar metallicity. Compared to Fig~\ref{411}, there is an increased number of detached convective layers with metallicity. For the 3000~K case, the increase of the convection layer height is larger with an increase in metallicity. There is also formation of two detached convective layers for a particle number per dry airmass of 10$^{7}$~kg$^{-1}$ and 10$^{8}$~kg$^{-1}$, though hardly discernible in the figure. There are three detached convective layers for the 4000~K case and four for the 5000~K. The height of these detached layers also vary. For the 5000~K, the detached layer can start as high as 7~10$^4$~Pa. A convective layer at this altitude could affect species condensing at lower temperature, such as Na$_2$S, ZnS or KCl. 

With a metallicity 10 times superior, Fe clouds with particle number density of 10$^9$ and 10$^{10}$~kg$^{-1}$ will trigger detached convective layer for the 5000~K case. On the other hand, the increase of metallicity for Al$_2$O$_3$ clouds does not trigger detached convective layers.
\bigbreak
For the CaTiO$_3$, Cr and MnS clouds, an increase of the metallicity by a factor 10 is far too small for these three clouds composition to have a significant impact. An increase of 100 will get cloud opacity greater than 1 for CaTiO$_3$ and Cr. 

\section{Effect on the emission spectra}
\label{Sec:disc}

\begin{figure}[H]
 \center
 \includegraphics[width=12cm]{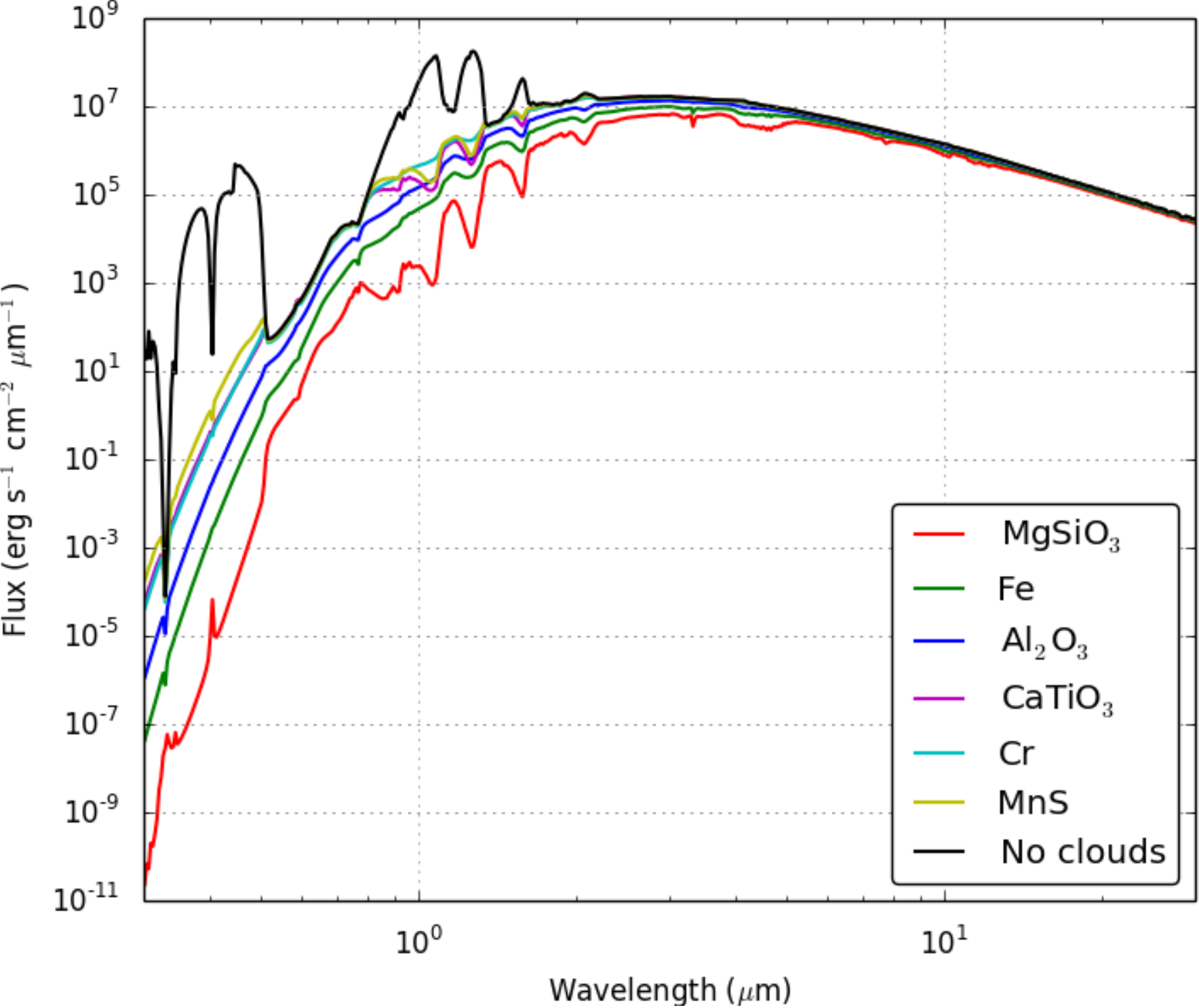}
 \caption{Thermal flux spectra for the six cloud composition considered with a cloud particle number density of 10$^8$~kg$^{-1}$ for the 4000~K case.}
 \label{51}
\end{figure}

We produce emission spectra of our model output using the 3D multiple-scattering Monte Carlo radiative-transfer code \textsc{gCMCRT} \citep{Lee21}. 1D averaged vertical profiles of the temperature, pressure and cloud properties across the simulation domain are produced for processing by \textsc{gCMCRT}. For chemical abundances, we assume chemical equilibrium and interpolate to a 2D table in temperature and pressure of species VMR produced using the \textsc{GGchem} code \citep{Woit18}, assuming solar metallicity elemental ratios from \citet{Aspl09}. These chemical abundance profiles are then used by \textsc{gCMCRT} to produce the opacity structure of the atmosphere. \textsc{gCMCRT} uses correlated-k tables for the gas opacity calculation; we produce k-tables consisting of 512 wavelength points between 0.3-30 $\mu$m (R $\approx$ 100), suitable for the resolution of the JWST NIRSPEC and MIRI instruments. K-tables are produced using cross-sections calculated using the HELIOS-K opacity code \citep{Grim15, Grim21}, using the following line-lists for different species: Na, K \citep{Kuru95}, H$_{2}$O \citep{Poly18}, CH$_{4}$ \citep{Harg20}, CO \citep{Li15b}, CO$_{2}$ \citep{Yurc20} and NH$_{3}$ \citet{Cole19}. K-coefficients in each band are weighted by the local VMR of spectrally active species using the random overlap method \citep[e.g.][]{Amun17}. \textsc{gCMCRT} also calculates the contribution from H$_2$, He and H CIA pairs \citep{Kar19} and Rayleigh scattering opacity sources. For the cloud opacity calculations, we assume the same size-distribution as used in the simulations, as well as the same MieX code \citep{Wolf04} and optical constants \citep{Kitz18}.
\bigbreak
Fig~\ref{51} shows the emission spectra for the six cloud composition considered with a cloud particle density of 10$^8$~kg$^{-1}$ for the 4000~K case, corresponding to the Fig~\ref{401}. As expected, the species that have the most impact on the flux are the one with the most impact on the deep convection, i.e. MgSiO$_3$ and Fe and Al$_2$O$_3$. However, there are several orders of magnitude decrease of flux for each species, especially below 0.5~$\mu$m and between 1 and 2~$\mu$m. For the first one, the spectral features are flattened by the clouds, and for the second one the spectral features are flattened by some clouds, but there is also the generation of other spectral features. Above 6~$\mu$m there is no significant difference between the clear sky and clouds cases. The larger impact of the clouds is in the optical, whereas the brown dwarfs are preferably observed in the IR due to the low magnitude in the optical.

\begin{figure}[H]
 \center
 \includegraphics[width=17cm]{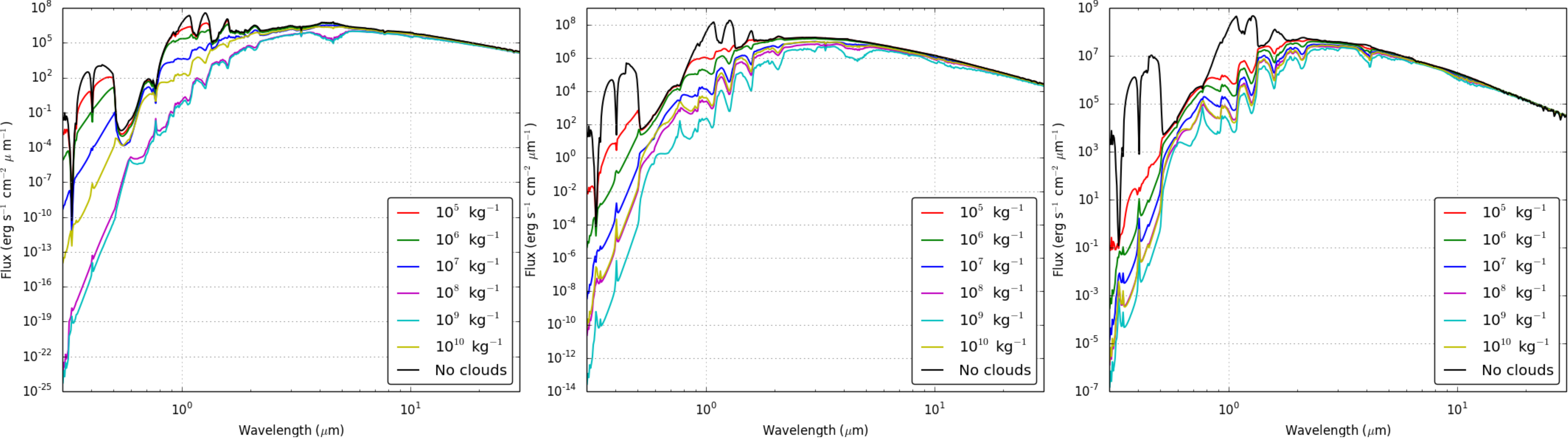}
 \caption{Thermal flux spectra with the MgSiO$_3$ for the 3000,4000 and 5000~K case and cloud particle number density between 10$^5$ and 10$^{10}$~kg$^{-1}$.}
 \label{52}
\end{figure}

Fig~\ref{52} shows the thermal flux spectra with MgSiO$_3$ cloud particle number density between 10$^5$ and 10$^{10}$~kg$^{-1}$ for the 3000, 4000 and 5000~K cases. For the 3000~K case, there are three groups, for 10$^5$ and 10$^6$~kg$^{-1}$ the flux decrease is small, for 10$^7$ and 10$^{10}$~kg$^{-1}$ the flux decrease is moderate and for 10$^8$ and 10$^9$~kg$^{-1}$ the flux decrease is strong, all linked to convective layer depth. The smoothing of the spectral features is strong for the 10$^8$ and 10$^9$~kg$^{-1}$ cases, with a new spectral feature between 4 and 5~$\mu$m. The trend in particle number density is similar for the two other temperature cases, the deeper the convective layer is the lower the thermal flux is. The smoothing of the spectral features below 0.5~$\mu$m is stronger with temperature, as for the spectral features between 1 and 2~$\mu$m.

\begin{figure}[H]
 \center
 \includegraphics[width=17cm]{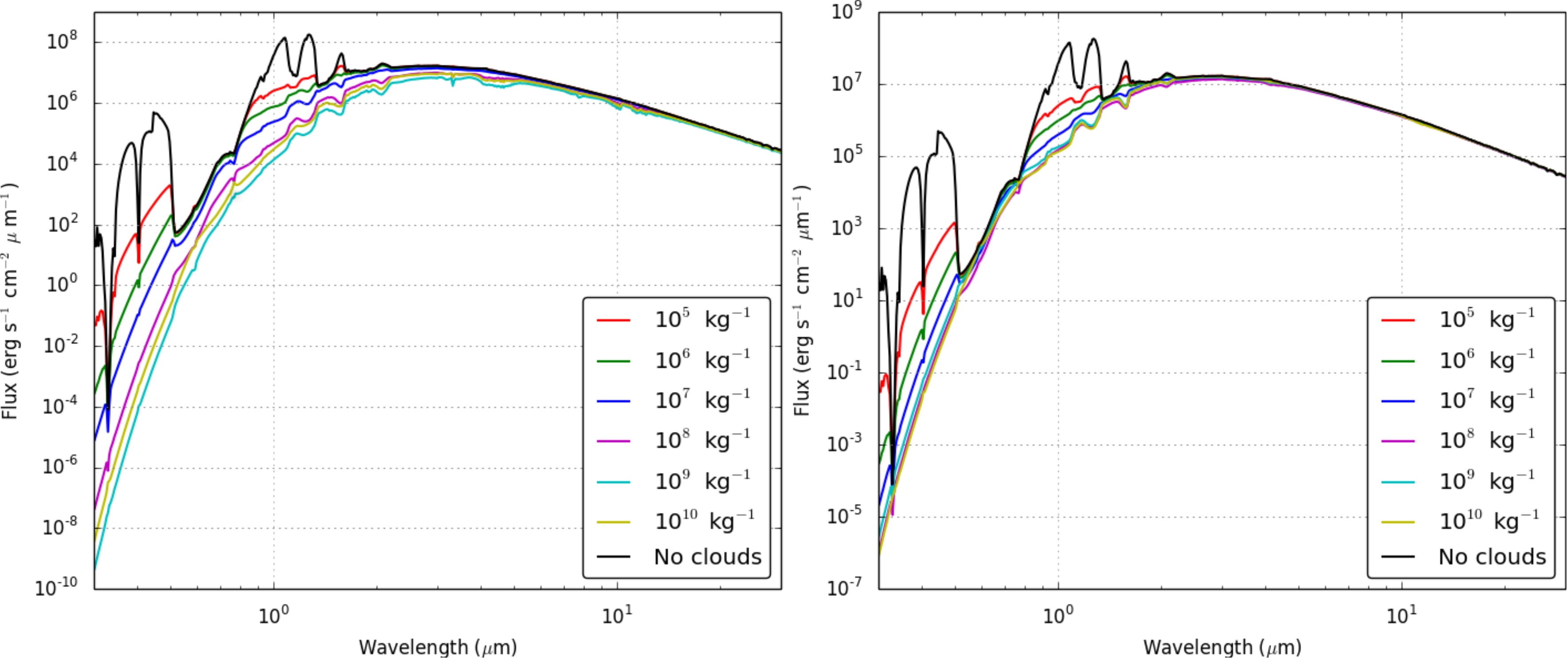}
 \caption{Thermal flux spectra with Fe and Al$_2$O$_3$ for the 4000~K case and cloud particle number density between 10$^5$ and 10$^{10}$~kg$^{-1}$.}
 \label{53}
\end{figure}

Fig~\ref{53} shows the same plot as Fig~\ref{51} but for Fe and Al$_2$O$_3$ clouds for the 4000~K case. As shown in Fig~\ref{421}, the impact of Fe on the height of the convective layer is stronger than the one of Al2O$_3$, and therefore the decrease of the thermal amplitude is stronger for Fe clouds. The cloud particle number density cases for which the increase of the convection depth is stronger, 10$^{9}$ for Fe and 10$^{10}$~kg$^{-1}$ for Al2O$_3$, have the strongest thermal flux decrease.

\begin{figure}[H]
 \center
 \includegraphics[width=12cm]{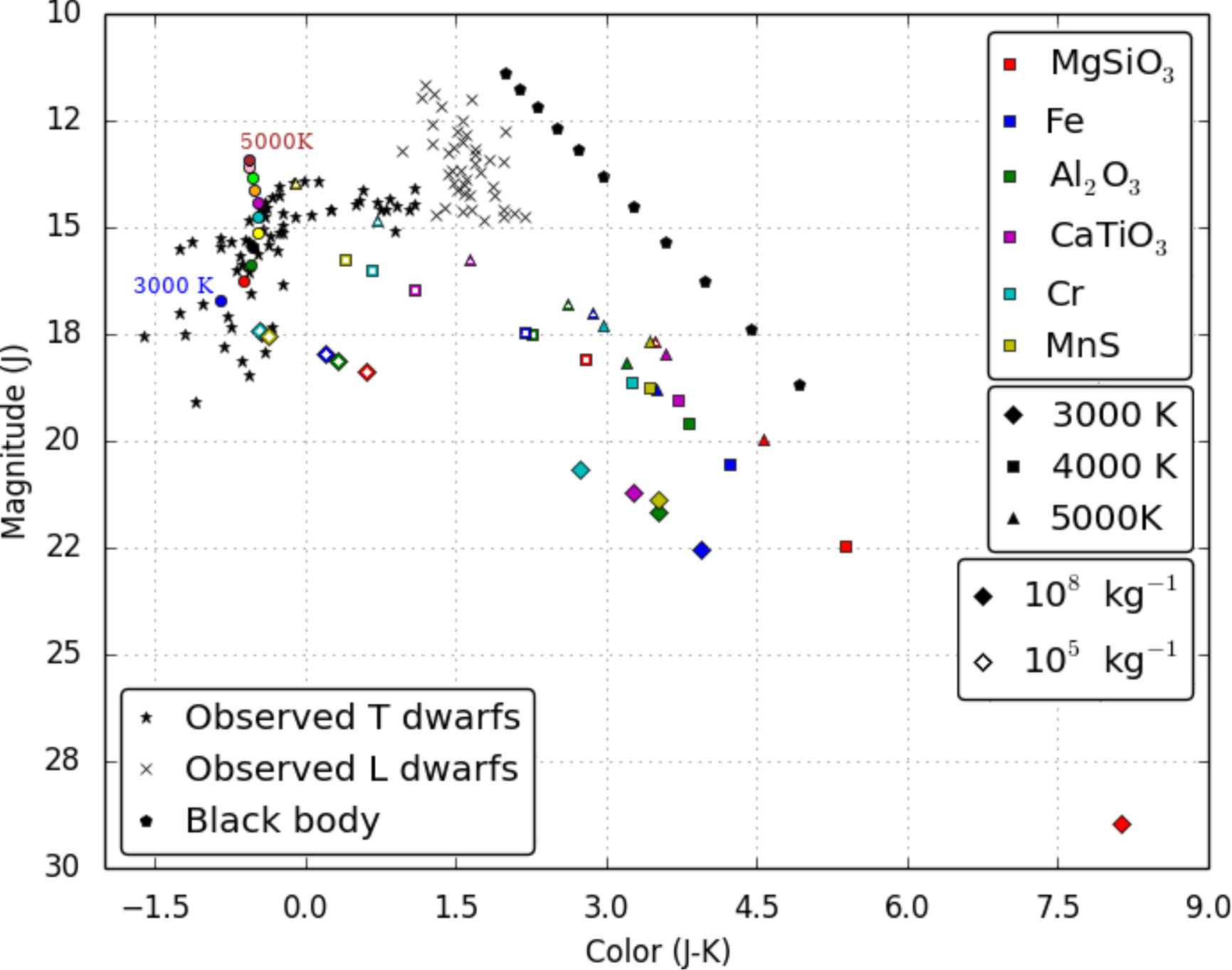}
 \caption{Magnitude-color diagram in the J and K band for the clear sky cases (circle, see Fig~\ref{241} for color bar), the six cloud species considered (see Fig~\ref{401} for color bar) for three temperatures and two cloud particle number density, 10$^5$ (empty symbols) and 10$^{8}$~kg$^{-1}$ (filled symbols), compared to observed brown dwarfs \citep{Dupu12}.}
 \label{54}
\end{figure}

Fig~\ref{54} shows the magnitude-color diagram in the J and K band for the clear sky and the different cloud considered compared to observed brown dwarfs \citep{Dupu12}. As expected, there is a strong effect of the clouds, with the clear sky points (circle) are being bluer than the clouds cases. Colors of the coldest modeled temperatures are consistent with some observed T dwarfs. The cloud cases have a J magnitude consistent with observed T dwarfs, but with much redder J-K color. There is a strong effect of the cloud particle number density, shown here with the 10$^5$ (empty symbols) and 10$^{8}$~kg$^{-1}$ (filled symbols). As shown in Section~\ref{Sec:cloud1}, a low value of particle number density will engender substantial cloud holes and the J-K color will be bluer. The 10$^{8}$~kg$^{-1}$ cases for MgSiO$_3$ exhibits very high J-K color, indicating that this value maybe not realistic as it is leading to cloud deck that is too thick and homogeneous. These simulations are performed with solar metallicity; higher metallicity will increase the depth of the convective layer and therefore the J-K color would be redder, but on the other hand with a lower metallicity the J-K color will appear bluer and closer to the observations. Our model does not represent large-scale features of the general circulation that could modulate cloudiness through large-scale uplift or subsidence.

\section{Conclusion}
\label{Sec:conc}
The model presented here is the first 3D convection-resolving model to model the radiative impact of the different types of cloud in brown dwarf atmospheres.\\
Without clouds, the convection depth increases with temperature, leading to larger convective cells and stronger vertical wind amplitude, reaching 600 m~s$^{-1}$ for the most extreme case. The vertical eddy diffusion from the resolved convection also increases with temperature, with values higher but consistent with previous estimation from the convection-resolving model and consistent with previous mixing-length theory estimations.\\
Several clouds compositions were tested with this model. The silicate clouds with MgSiO$_3$ are the one with the most impact. The depth of the convective region is increased by the heating at the cloud base, and for particle radius around 1~$\mu$m, this heating is so strong that it destabilizes the atmosphere leading to an independent detached cloud layer. The detached cloud is present for larger particle radius range as the temperature increases. The temporal frequency is of similar order to \cite{Tan19} 1D model, however the domain-mean variability is significantly weaker.\\
The cloud particle number per dry airmass has a significant impact on the presence of cloud holes. For values above 10$^{7}$~kg$^{-1}$ there is a complete cloud coverage. For values below, there is a significant presence of cloud holes. The temperature also has an impact on the presence of cloud holes; at low temperature the clouds are puffier, whereas at higher temperature there is an aggregation of clouds due to the size of the convective cells.\\
The Fe and Al$_2$O$_3$ clouds also have a significant impact on the deep convective layer depth, but the heating at cloud base is not strong enough to engender a detached cloud convective layer.\\
However, cloud particle of CaTiO$_3$, Cr and MnS have limited impact on the convection depth and therefore on the emission spectra, due to the low abundance of condensable gas and the thin layer where the cloud condenses.\\
Two metallicities have been used in this study, a solar metallicity and ten times the solar metallicity. The depth of the clear sky convective layer is slightly deeper, with an increased metallicity. The depth of the detached cloud convective layer is more affected, with a significant increase.\\
The different cloud composition affects more or less the deep convection depth and the thermal structure above, and therefore impacts the thermal spectra. There is a decrease of the flux amplitude and a smoothing of the spectra due to the clouds. MgSiO$_3$ clouds have the most impact on the emission spectra. The effect of $\sigma$ (the nondimensionalized constant measuring the width of the particle size distribution) on the cloud deck was not studied, however it could have significant effect on the effective radius and therefore on the radiative properties of the clouds.\\
This cloud-convective feedback could also be present in hot giant exoplanets; the insolation and the specific star spectra could have a strong effect on the cloud layer. The insolation could bring extra heat and destabilize the atmosphere like for Earth, or on the contrary stabilize a region like Venus deep cloud \citep{Imam14,Lefe18}. Different spectra could have an impact on the energy deposition in the atmosphere, as for Proxima Centauri-b \citep{Lefe21}. In hot jupiters cloud studies, the moment formalism is sometimes used in the microphysical modelling \citep{Lee16,Line18}, allowing the particle number density to be then determine, as well as different particle size distribution \cite{Chri21}. 
\bigbreak
\cite{Char18} showed a strong gravity effect on the chemistry and microphysics and the overall color. A similar study needs to be conducted with both higher and lower gravity to capture its impact on the cloud deck altitude and depth.\\
A non-grey radiative transfer scheme would be beneficial to fully understand the radiative effects of the different cloud compositions, with the correlated-k formalism for example. The feedback due to specific absorption/emission lines, with a spectral dependence of the extinction and scattering coefficients for the different cloud species, could be therefore quantified. \\
In this study, no wind shear was imposed on the domain and the depth of the deep convective layer is governed by the IR heating of the deep atmosphere. However, the dynamics and global atmospheric circulation will impose wind shear that could affect convection as well as gravity waves generation and vertical propagation \citep{Frit03}. Such gravity waves could impact the condensation and nucleation of species condensing at higher altitudes \citep{Pare20}.\\
The heat capacity of H/He is set constant in the model, while the temperature varies by several thousands of kelvin over the vertical. In such temperature range, the heat capacity variation should be taken into account in future work as it could impact the stability in parts of the atmosphere.\\
The condensation process in our model is highly simplified, with silicate vapour, for example, condensing into enstatite; this is an approximation and a chemical model is therefore needed to resolve the complexity of such environments. Such improvement will permit the simultaneous presence of multiple cloud types and the interaction between the different kinds of clouds.\\
The microphysical scheme used in this study is idealized, coagulation and growth of cloud particle are not taken into account and effects of non-spherical particles, micro-porosity \citep{Samr20} and resolving the cloud particle size distribution \citep[e.g.][]{Gao18} are not implemented.
The latent heat release of the different clouds is overlooked in the study, and an implementation of such energy release could help the study of local storms \citep{Tan17}.

Convective-resolving studies have been used for Earth \citep{Rio10} and Mars \citep{Cola13} atmosphere to improve the parametrization of the convection in the GCMs. The present study could be the starting point of such methodology for brown dwarfs and hot giant exoplanets.

\section*{Acknowledgements}
The authors would like to thank the anonymous reviewer for the helpful comments that improved the overall quality of the paper.
This project has received funding from the European Research Council (ERC) under the European Union’s Horizon 2020 research and innovation program (grant agreement No. 740963/EXOCONDENSE). The authors would like to acknowledge the use of the University of Oxford Advanced Research Computing (ARC) facility in carrying out this work. \url{http://dx.doi.org/10.5281/zenodo.22558}. The Rosseland weighting cloud opacity code can be found on GitHub: \url{https://github.com/ELeeAstro/Rosseland_Clouds}

\appendix
\begin{appendices}
\section{Rosseland Mean}
\label{App}

\begin{figure}[H]
 \center

Fig~\ref{A11} shows the Rosseland mean extinction coefficient for the six cloud species considered over the temperature and particle radius range. The extinction coefficient have a similar structure for all the species, with a value below one for radii below 1$\mu$m and a value above 1 for radii above 1$\mu$m. The transition radius varies with the species and temperature. MgSiO$_3$ particles have the highest transition radius, whereas Cr particles have the lowest. Cr and Fe particles have a temperature dependence of the extinction coefficient between 1$\mu$m and 10$\mu$m, where the coefficient is stronger at high temperatures. \\
 
 \includegraphics[width=14cm]{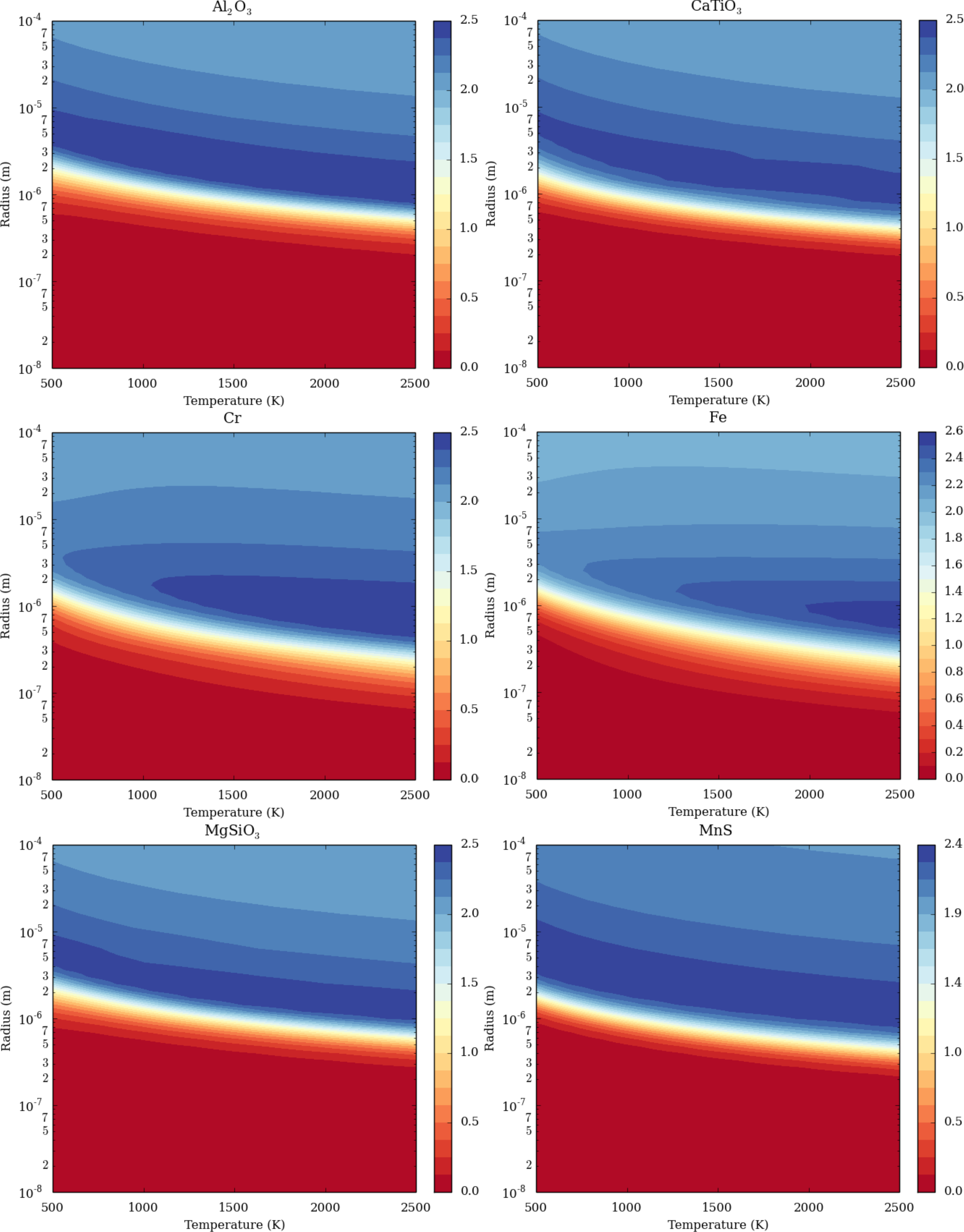}
 \caption{Rosseland mean extinction coefficient Q$_{ext}$ for Al$_2$O$_3$, CaTiO$_3$, Cr, Fe, MgSiO$_3$, MnS clouds.}
 \label{A11}
\end{figure}

\begin{figure}[H]
 \center
 \includegraphics[width=14cm]{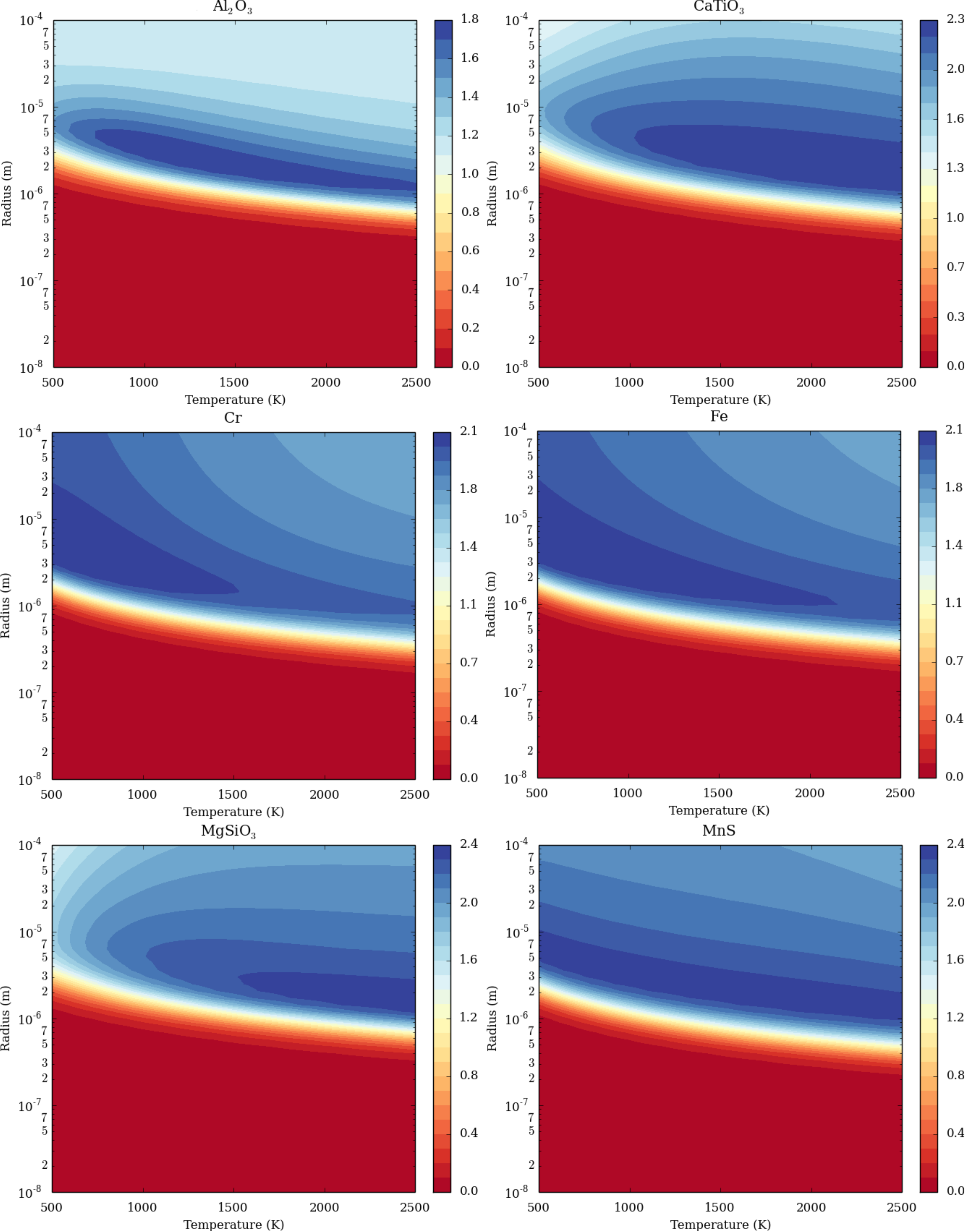}
 \caption{Rosseland mean single-scatter coefficient Q$_{scat}$ for Al$_2$O$_3$, CaTiO$_3$, Cr, Fe, MgSiO$_3$, MnS clouds.}
 \label{A12}
\end{figure}

Fig~\ref{A12} shows the Rosseland mean single-scatter coefficient for the six cloud species considered over the temperature and particle radius range. The scattering coefficient have a similar structure for all the species, with a value below one for radii below 1$\mu$m and a value above 1 for radii above 1$\mu$m. The transition radius varies with the species and temperature. MgSiO$_3$ particles have the highest transition radius, whereas Cr particles have the lowest. Contrary to the extinction coefficient, all the species have a temperature dependence for radius above 1$\mu$m. Al$_2$O$_3$, Cr, Fr and Mns have a stronger coefficient for lower temperature whereas for CaTiO$_3$ and MgSiO$_3$ it is the opposite.\\

\begin{figure}[H]
 \center
 \includegraphics[width=14cm]{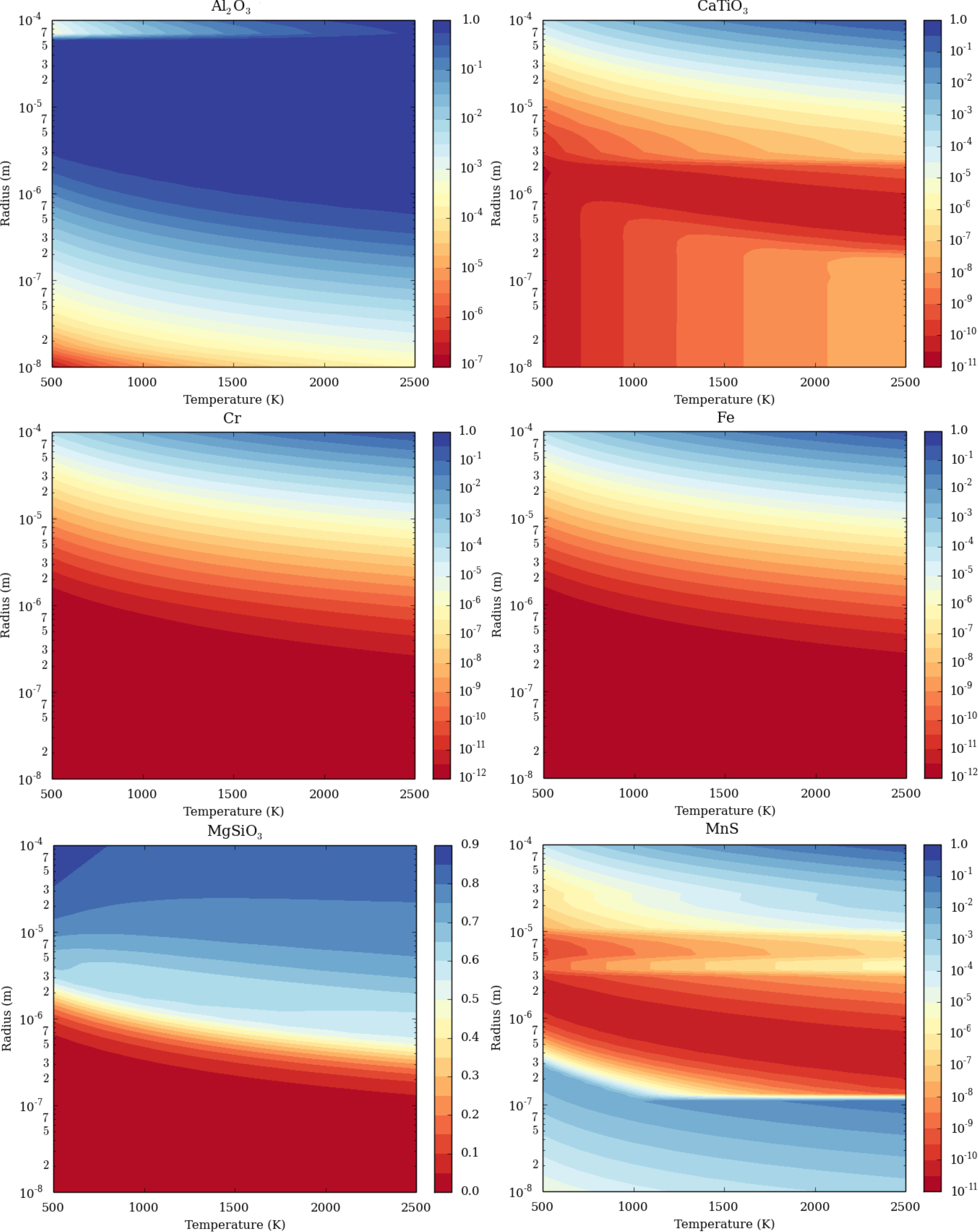}
 \caption{Rosseland mean asymmetry factor for Al$_2$O$_3$, CaTiO$_3$, Cr, Fe, MgSiO$_3$, MnS clouds.}
 \label{A13}
\end{figure}

Fig~\ref{A13} shows the Rosseland mean asymmetry factor for the six cloud species considered over the temperature and particle radius range. Contrary to extinction and scattering coefficient, there is a broad range of values and radius dependency for the asymmetry factor. Al2O3 and MgSiO3 particles have values above 0.1 above radius of 1$\mu$m, with a drop for radius close to 100$\mu$m. Cr and Fe particles have an asymmetry factor varying from 10$^{-12}$ for radius 1$\mu$m and below to 1 for a 100$\mu$m radius at high temperature. CaTiO3 particles have an asymmetry factor varying from 10$^{-7}$ for the lowest radius to 0.3 for the highest, with a strong drop for radius around 100$\mu$m. MnS have an asymmetry factor between 10$^{-3}$ and 10$^{-2}$ for all the radius range, with a strong drop for radius between 1$\mu$m to 10$\mu$m.

\newpage
\section{MgSiO$_3$ clouds}
\label{App2}
\begin{figure}[H]
 \center
 \includegraphics[width=17cm]{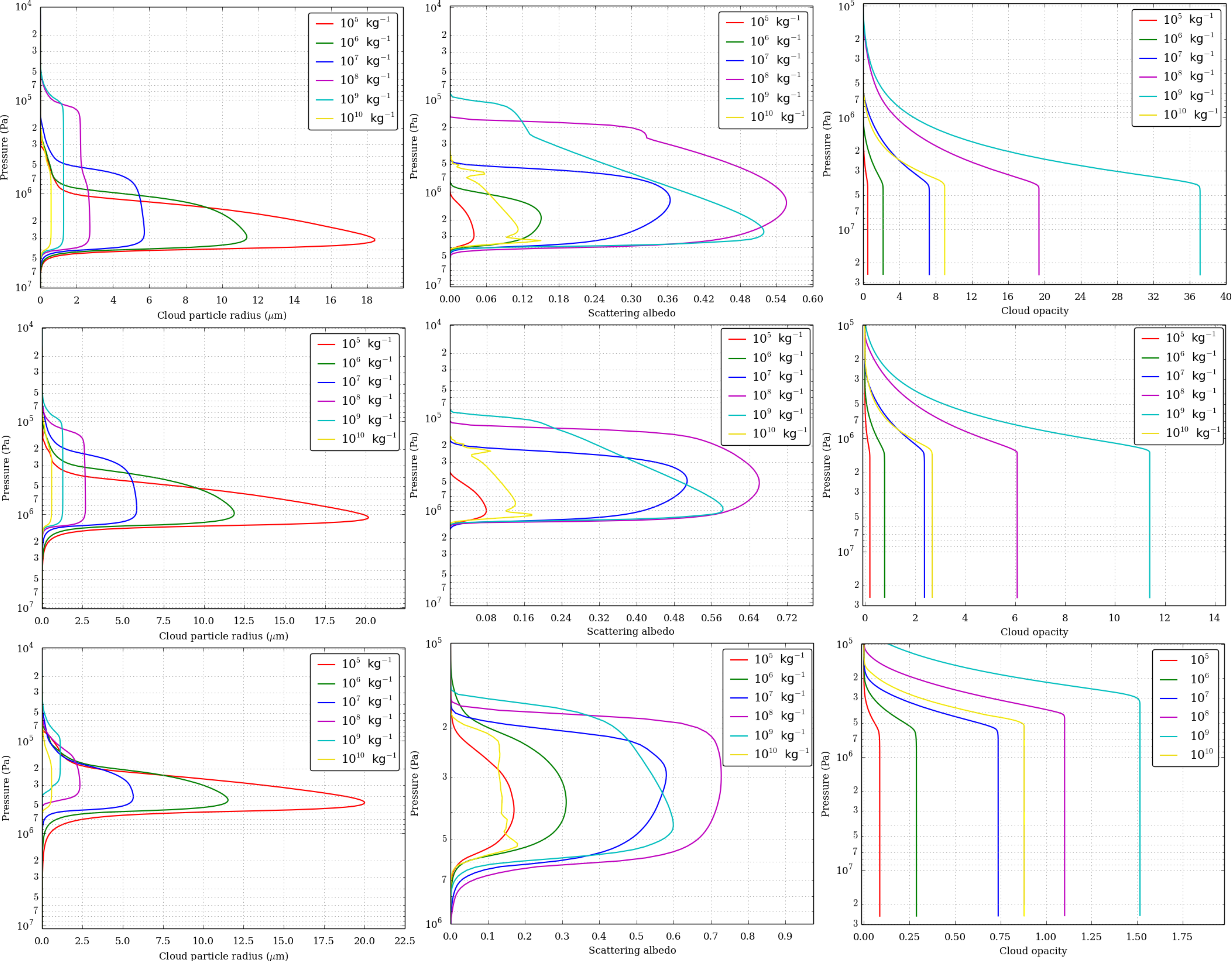}
 \caption{Domain averaged vertical profiles of the reference radius (left column), Scattering albedo (middle column) and cloud opacity (right column) for the 3000~K (top row), 4000~K (middle row) and 5000~K (bottom row) case with MgSiO$_3$ clouds and particle number per dry airmass between 10$^5$ and 10$^{10}$~kg$^{-1}$.}
 \label{A21}
\end{figure}

\section{Fe and Al2O$_3$ clouds}
\label{App3}
\begin{figure}[H]
 \center
 \includegraphics[width=17cm]{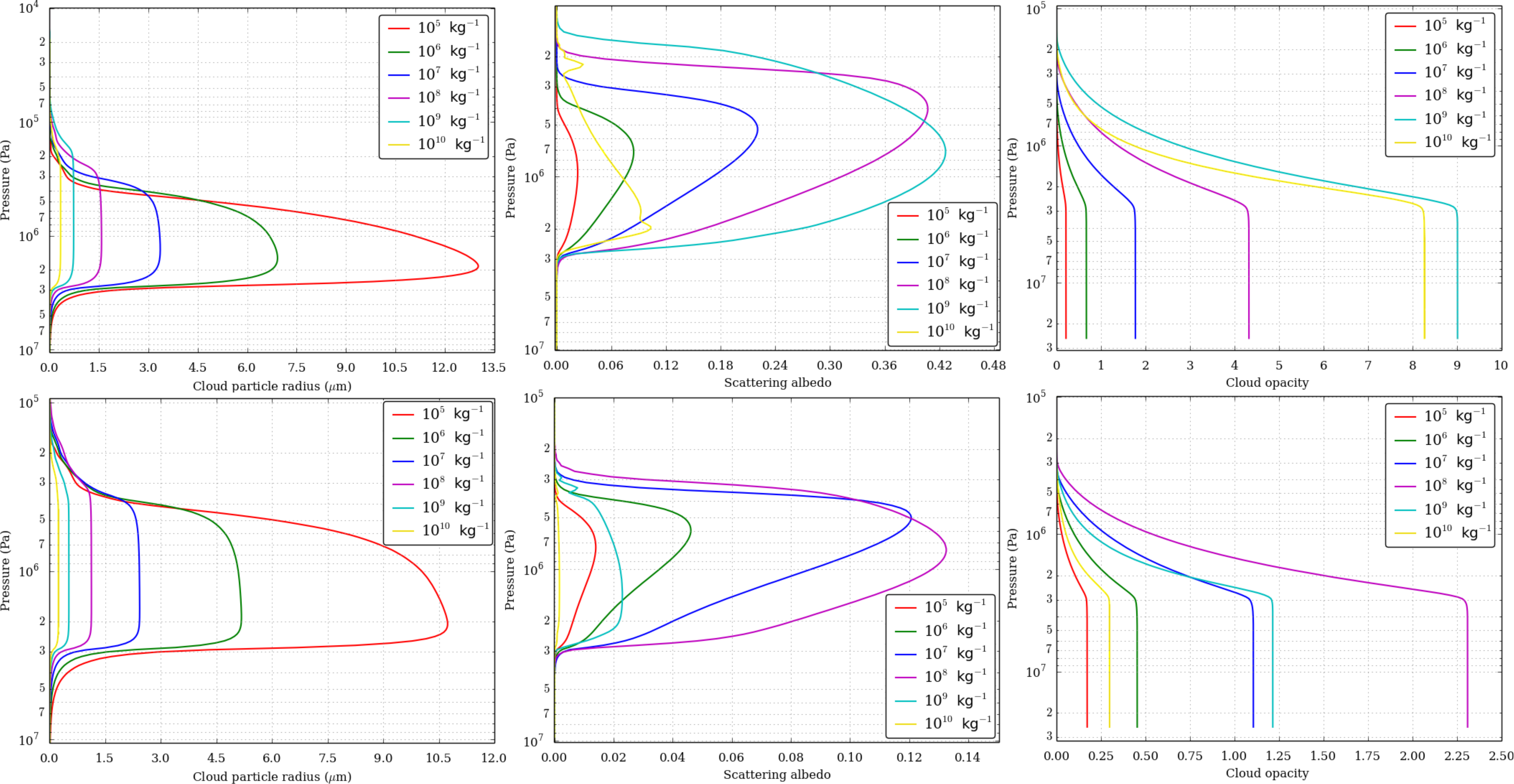}
 \caption{Domain averaged vertical profiles of the reference radius (left column), Scattering albedo (middle column) and cloud opacity (right column) for the 4000~K case with Fe (top row) and Al$_2$O$_3$ (bottom row) clouds and particle number per dry airmass between 10$^5$ and 10$^{10}$~kg$^{-1}$.}
 \label{A31}
\end{figure}

\end{appendices}

\end{document}